\documentclass[tightenlines,aps,floatfix,preprint,nofootinbib]{revtex4}
\usepackage{float,axodraw,epsfig}

\newcommand{\HWW}{$\rm{gg \rightarrow H \rightarrow WW \rightarrow \ell \nu \ell \nu}$}

\newcommand{\mr}[1] {\mathrm{#1}}

\newcommand{\pt} {\ifmmode p_\mathrm{T}\else   $p_\mathrm{T}$\fi}
\newcommand{\ptH} {\ifmmode p_\mathrm{T}^\mathrm{H}\else   $p_\mathrm{T}^\mathrm{H}$\fi}
\newcommand{\ptWW} {\ifmmode p_\mathrm{T}^\mathrm{WW}\else   $p_\mathrm{T}^\mathrm{WW}$\fi}
\newcommand{\ptJ} {\ifmmode p_\mathrm{T}^\mathrm{Jet}\else   $p_\mathrm{T}^\mathrm{Jet}$\fi}
\newcommand{\ptmin} {\ifmmode p^\ell_\mathrm{T min}\else   $p^\ell_\mathrm{T min}$\fi}
\newcommand{\ptmax} {\ifmmode p^\ell_\mathrm{T max}\else   $p^\ell_\mathrm{T max}$\fi}
\newcommand{\fl}{\rightarrow}

\newcommand{\newc}[1] {\newcommand{#1}}
\newc{\R}{$R$}
\newc{\charginom}{M_{\tilde \chi}^{+}}
\newc{\mue}{\mu_{\tilde{e}_{iL}}}
\newc{\mud}{\mu_{\tilde{d}_{jL}}}
\newc{\barr}{\begin{eqnarray}}
\newc{\earr}{\end{eqnarray}}
\newc{\beq}{\begin{equation}}
\newc{\eeq}{\end{equation}}
\newc{\ra}{\rightarrow}
\newc{\lam}{\lambda}
\newc{\eps}{\epsilon}
\newc{\gev}{\,GeV}
\newc{\tev}{\,TeV}
\newc{\eq}[1]{(\ref{eq:#1})}
\newc{\eqs}[2]{(\ref{eq:#1},\ref{eq:#2})}
\newc{\etal}{{\it et al.}\ }
\newc{\eg}{{\it e.g.}\ }
\newc{\ie}{{\it i.e.}\ }
\newc{\nonum}{\nonumber}
\newc{\lab}[1]{\label{eq:#1}}
\newc{\dpr}[2]{({#1}\cdot{#2})}
\newc{\gsim}{\stackrel{>}{\sim}}
\newc{\lsim}{\stackrel{<}{\sim}}

\newc{\gevcc}{GeV}
\newc{\tevcc}{TeV}
\newc{\gevc}{GeV}
\newc{\PY}{\textsc{PYTHIA}}

\newcommand{\ba}{\begin{eqnarray}}
\newcommand{\ea}{\end{eqnarray}}

\newcommand{\be}{\begin{equation}}
\newcommand{\ee}{\end{equation}}
\def\eq#1{(\ref{#1})}

\def\hname{{\sf ~FEHiP}}
\begin{document}
\begin{titlepage}
\begin{flushright}
{ETHZ-IPP  PR-06-05} \\
{April 16, 2007}\\
\end{flushright}

\begin{center}

{\bf \LARGE
Simulation of a Cross Section and Mass \\ \smallskip Measurement of a SM Higgs Boson in the \\ \smallskip \smallskip \smallskip
\HWW \ Channel at the LHC
}
\end{center}

\smallskip \smallskip 
\begin{center}
{\Large G. Davatz, M. Dittmar and F. Pauss}
\end{center}
\bigskip

\begin{center}
Institute for Particle Physics (IPP), ETH Z\"{u}rich, \\
CH-8093 Z\"{u}rich, Switzerland
\end{center}

\smallskip 
\begin{abstract}

\smallskip \smallskip

The potential to discover a Standard-Model-like Higgs boson at the LHC in the mass range from 150-180 GeV, decaying into a pair of W bosons with subsequent leptonic decays, has been established during the last 10 years.
Assuming that such a signal will eventually be observed, the analysis described in this paper  
investigates how accurate the signal cross section can be measured and how the observable lepton \pt\ spectra can be used to constrain the mass of the Higgs boson.
Combining the signal cross section with the analysis of the lepton \pt\ spectra and assuming the SM Higgs cross section is known with an accuracy of $\pm$ 5\%, our study indicates that an integrated luminosity of about 10 fb$^{-1}$ allows to measure 
the mass of a SM Higgs boson with an accuracy between 2 and 2.5 GeV.

\vspace{2cm}
\end{abstract}

\smallskip \smallskip 
\begin{center}
{\it submitted to Physical Review D }
\end{center}
\end{titlepage}

\maketitle

\section{Introduction}
The search for a Standard-Model(SM)-like Higgs boson in the mass range between the current limit 
of about 115 GeV and about 1 TeV is one of the main goals of ATLAS and CMS, the two large LHC experiments~\cite{search}.
A large variety of signatures have been studied in detail during the last 15 years and the main discovery channels are now 
well established. For a detailed description of all these possibilities, we refer the reader to Refs.~\cite{search} and some recent 
reviews~\cite{jakobs,djouadi_rev,carena}. 

The three main Higgs discovery channels are the $\rm{\gamma\gamma}$ channel ($\rm{gg \to H \to \gamma \gamma}$) for masses below 140 GeV, the four lepton channel ($\rm{gg \to H \to ZZ^{(*)}\to \ell \ell \ell \ell}$)  
for masses between 130-155 GeV and above 180 GeV, and the $\rm{H \to WW}$ channel (\HWW) for masses between 150 GeV to 180 GeV~\cite{Dittmar:1996ss}.

Signals in the four lepton channel and the $\rm{\gamma\gamma}$ channel would be observed as peaks in the invariant mass 
above backgrounds. These peaks are unambiguous discovery signals, which can be used directly for mass and cross section measurements.
Such measurements are essentially only limited by statistics. It has been estimated that these mass peaks, once a luminosity 
between 10-100 fb$^{-1}$ is accumulated, can be used to measure the Higgs mass directly with an accuracy smaller than 
1 GeV. A detailed description of these possibilities can be found in some recent 
reviews~\cite{jakobs,djouadi_rev,carena}. 

The $\rm{H \to WW}$ channel with the subsequent W decays to leptons ($\rm{W \to}\ \ell \nu$) is now well 
established as the discovery channel in the Higgs mass region between 150 GeV to 180 GeV.
In contrast to the other two channels, the two undetectable neutrinos do not allow the direct reconstruction of a 
mass peak and some model dependent Monte Carlo techniques have to be used.

As this mass region is otherwise almost inaccessible and the proposed signature has a large signal cross section 
with a relative good signal to background ratio, it is interesting to investigate 
how and how well the Higgs mass can be determined in this channel with two undetectable neutrinos.

Such procedures are not new and have previously been used successfully for the mass determination of the 
W boson and the top quark at hadron colliders (see for example Ref.~\cite{genzinski}). 

So far only qualitative procedures have been proposed, using either the charged lepton \pt\ spectra, as suggested in Ref.~\cite{dittmareps97}, or the transverse mass distribution, as proposed in Refs.~\cite{Glover:1988fn, Rainwater:1999sd}.

The transverse mass determination involves directly the measurement of the
missing transverse momentum with the related experimental uncertainty and depends to some extend also on the various
required selection cuts. The missing transverse momentum can be defined in several ways, and depends on many details of the actual detector behavior. A detailed analysis of the corresponding systematic errors of this observable has not been performed yet. 
In contrast, the mass-dependent observables studied in this paper are much simpler and the systematic uncertainties can already be determined. 

This analysis, described in the following, uses the observed signal cross section for the $\rm{H \to WW}$ channel for a Higgs mass between 150-180 GeV
and the observable lepton \pt\ spectra.
For this study we use the reweighting method, described in Ref.~\cite{Davatz:2004zg}, 
which allows to approximate higher order QCD corrections to the Higgs \pt\ spectrum. 
A similar study was performed in Ref.~\cite{Davatz:2006ut} including also rapidity-dependence by 
reweighting to FEHIP\cite{Anastasiou:2005qj}, a program computing fully differential distributions for Higgs production at NNLO. 
It was shown there that the effect of the rapidity-dependence is very small. To simplify the analysis the signal selection criteria 
from Ref.~\cite{Davatz:2004zg}, which are based on the criteria proposed in 
Refs.~\cite{Dittmar:1996ss} and \cite{dittmareps97}, are used.

This paper is structured as follows. First (Section II) we repeat the analysis described in Ref.~\cite{Davatz:2004zg} using   
mass independent selection  
criteria.

Next (Section III) we analyze the ``observed'' signal cross section 
for different Higgs masses and within the Standard Model. 
The potential systematic uncertainties originating from the various backgrounds and with essentially identical selection criteria
have recently been determined in a full CMS detector simulation~\cite{Davatz:2006kb}.
The results from this analysis are used in the following to estimate the potential systematic errors for the mass measurement.

Next (Section IV) we investigate how correlations between the Higgs mass and the observable lepton \pt\ spectra can be used to constrain the Higgs mass.
Finally (Section V), it is demonstrated how the mass ambiguities, which result from the signal cross section measurement, can be 
resolved from a detailed analysis of the lepton \pt\ spectra. The uncertainties from 
the remaining model dependence are also discussed.

\section{Selection of Higgs signal events}
\label{sec:signalcuts}

The possibility to distinguish the Higgs boson in the decay $\rm{H \to WW \to \ell \nu \ell \nu}$ 
from non resonant backgrounds and for masses between 150-180 GeV is based 
on the following qualitative criteria: 

\begin{itemize} 
\item In contrast to the various backgrounds from non-resonant $\rm{WW}$ decays, the two charged leptons originating from the Higgs decay have a small opening angle.
This particular signal structure originates from the correlated spins of the two W bosons produced in the decay 
of a a spin 0 object and from the V-A structure of W decays. This behavior is preserved as long as the 
transverse momenta of the W's are small compared to the W mass. This condition is fulfilled if the Higgs mass is close to the mass of two W bosons
and if the Higgs boson itself is produced with a small transverse momentum.

\item The $\rm{WW}$ pair, originating from the Higgs decay, is produced dominantly in the gluon fusion process while the 
continuum $\rm{WW}$ events are produced mainly from $q\bar{q}$ scattering. As a result, the signal events have a shorter rapidity plateau  
then the continuum $\rm{WW}$ background events.
\item $\rm{WW}$ pairs originating from the production and decay of top quark pairs are usually accompanied by jets 
and can be strongly reduced by a properly adjusted jet veto.
\item The observable lepton transverse momentum spectra show a Jacobian peak-like structure
allowing to further optimize the signal over background ratio. 
\end{itemize}

The Higgs signal events are simulated with the PYTHIA Monte Carlo ~\cite{Sjostrand:2006za}, where the 
Higgs \pt\ spectrum is reweighted 
such that it matches the one expected from the resumed NNLO(+NNLL) calculations (generated with the HqT program, 
Ref.~\cite{Bozzi:2003jy,Bozzi:2005wk}). 

Some first optimization of the selection cuts for a Higgs mass around 165 GeV has been described previously ~\cite{Dittmar:1996ss, dittmareps97, Davatz:2004zg}.
For the purpose of this paper 
we apply the selection criteria used in Ref.~\cite{Davatz:2004zg}. 
The proposed analysis proceeds in two steps. First, criteria 1-5 selects events which contain two isolated 
high \pt\ leptons which come largely from resonant or non resonant events of the type $\rm{WWX}$. 
In the second group of cuts, criteria 6-8, the resonant $\rm{H \rightarrow WW}$ signal events are separated from continuum $\rm{WWX}$ events.

In detail, the following cuts are applied:
\begin{enumerate}
\item The event should contain two leptons, electrons or muons, with opposite charge, 
each with a minimal \pt\ of 10 \gevc\ and a 
pseudorapidity $|\eta|$ smaller than 2.
\item
In order to have isolated leptons, it is required that the 
transverse energy sum from detectable particles, defined as ``stable'' charged or neutral particles with a \pt\ larger than 
1 \gevc, found inside a cone of 
$\Delta R = \sqrt{\Delta \eta^2+\Delta \phi^2)}< 0.5$
around the lepton direction, should be smaller than 10\% of the
lepton energy and the invariant mass of all detectable particles within the 
cone should be smaller than 2 \gevcc. 
Furthermore, at most one additional detectable particle inside a cone of $\Delta R < 0.15$ is allowed.
\item
The dilepton mass, $m_{\ell \ell}$, has to be smaller than 80 \gevcc. 
\item
The missing \pt\ of the event, required to balance the 
\pt\ vector sum of the two leptons, should be larger than 20 \gevc.
\item
The two leptons should not be back-to-back in the plane 
transverse to the beam direction. The opening angle 
between the two leptons in  this plane is required to be smaller than 135$^{\circ}$. 
\end{enumerate}

Dilepton events, originating from the decays of W and Z bosons, 
are selected with the criteria 1 and 2. Lepton pairs
which originate from the inclusive production of $\mr{Z} \rightarrow \ell \ell (\gamma)$, 
including Z decays to $\rm{\tau}$ leptons, are mostly removed with criteria 3--5.
 
Starting with this initial set of requirements, the following criteria exploit the 
differences between Higgs events 
and the so-called ``irreducible" background from continuum production 
of $\mr{pp} \ra \mr{W}^{+} \mr{W}^{-}$ events \footnote{In a recent paper the importance of the heavy flavour background has been investigated \cite{Sullivan:2006hb}. As our selection criteria are much stronger than the ones discussed in this paper, no relevant contribution from this background is expected.}.
\begin{enumerate}
\setcounter{enumi}{5}

\item The opening angle $\phi$
between the two charged leptons in the plane transverse
to the beam should be smaller 
than 45$^{\circ}$
and the invariant mass of the lepton pair should be smaller 
than 35 \gevcc\footnote{A minimal angle of 10$^{\circ}$ (or minimal mass of 10 GeV) might be 
needed in order to reject badly measured $\Upsilon \rightarrow \mr{e}^{+}\mr{e}^{-} (\mu^{+}\mu^{-})$ decays.
Such a cut would not change the signal efficiency in any significant way.}.
\item
Jets are formed with a cone algorithm, requiring a minimum 
jet transverse momentum in order to be considered as a jet. For this analysis, events which contain a jet, with \ptJ\ $>$ 30 GeV and with a  
pseudorapidity $|\eta^\mr{Jet}|$ $<$ 4.5, are removed.

\item
Finally, the \pt\ spectrum of the two charged leptons is exploited. For this, the two leptons
are classified according to their \pt (\ptmin\ and \ptmax). 
It is found that the \ptmax\ and \ptmin\
show a Jacobian peak-like structure for the signal, 
which depends on the simulated Higgs mass. 
In case of a Higgs mass close to 165 \gevcc,  
\ptmax\ should be between 35 and 50 \gevc, whereas 
the \ptmin\ should be larger than 25 \gevc.

\end{enumerate}

\begin{figure}[!h]
\begin{center}
\begin{minipage}{\linewidth}
\includegraphics*[scale=0.4]{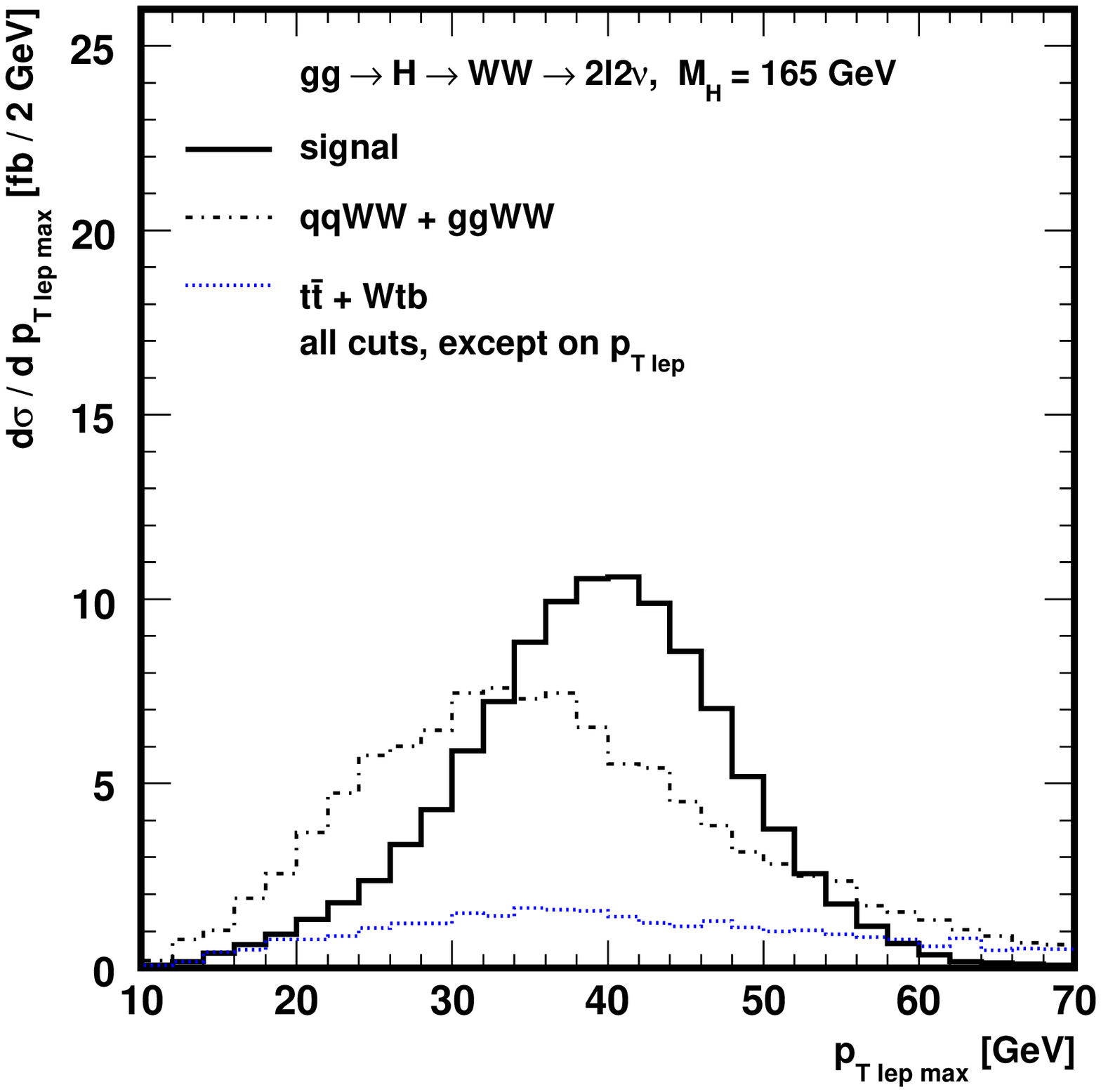}
\includegraphics*[scale=0.4]{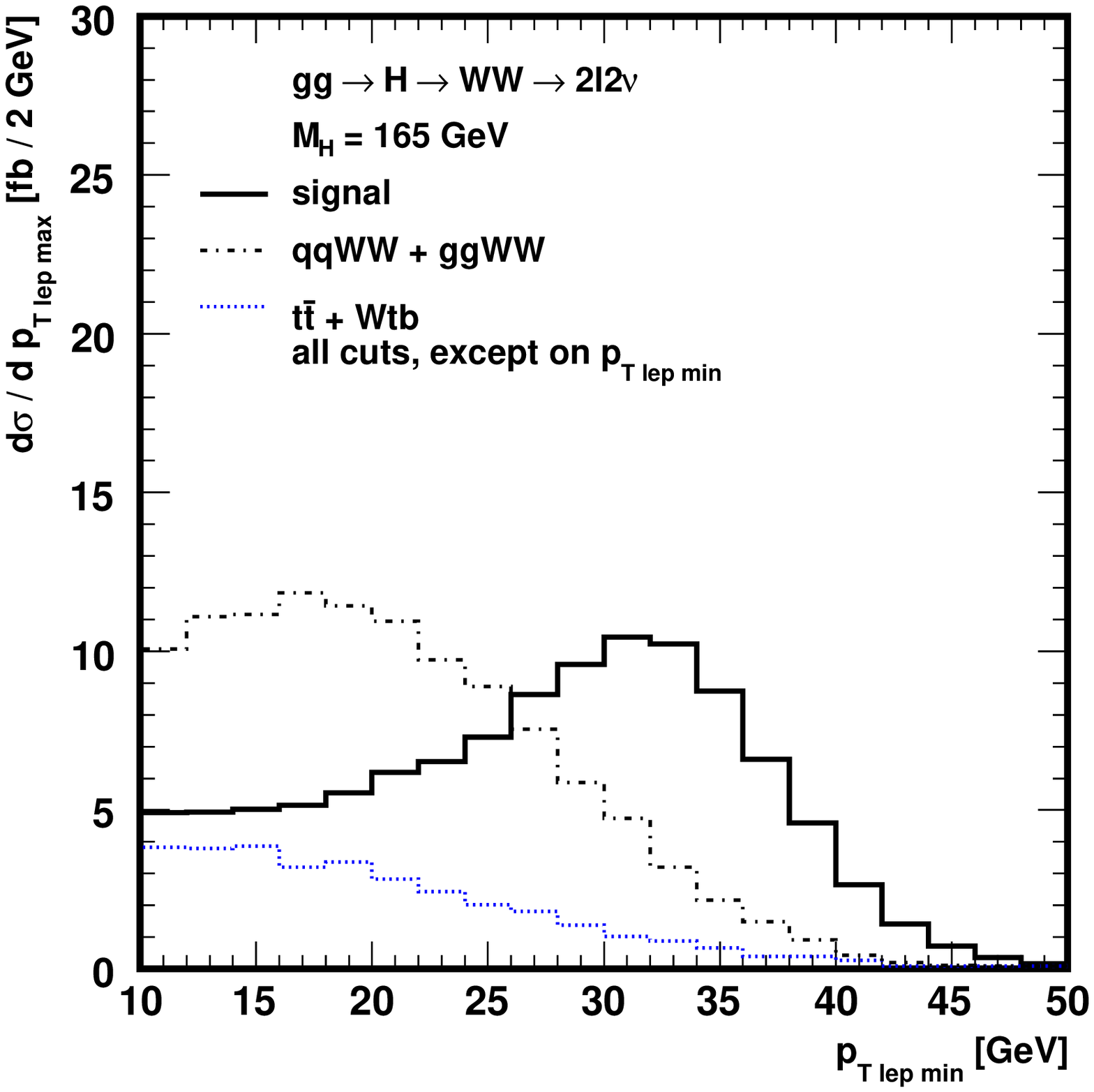}
\caption{Cross section of the Higgs signal ($\rm{M_H}$ = 165 GeV) as well as the main backgrounds, that is qqWW, ggWW, $\rm{ t\bar{t}}$ and Wtb, 
as a function of the \ptmax\ (left) and \ptmin\ (right). All cuts are applied except the 
ones on the \pt\ of the leptons. The events are generated with PYTHIA and TOPREX \cite{Slabospitsky:2002ag} and reweighted to NNLO and NLO respectively. WW
production via gluon fusion was generated using a Monte Carlo provided by N. Kauer \cite{Binoth:2005ua}, with parton shower
simulation in PYTHIA.}
\end{minipage}\label{p3}
\end{center}
\end{figure}
\begin{figure}[h!]
\begin{center}

\includegraphics*[scale=0.4]
{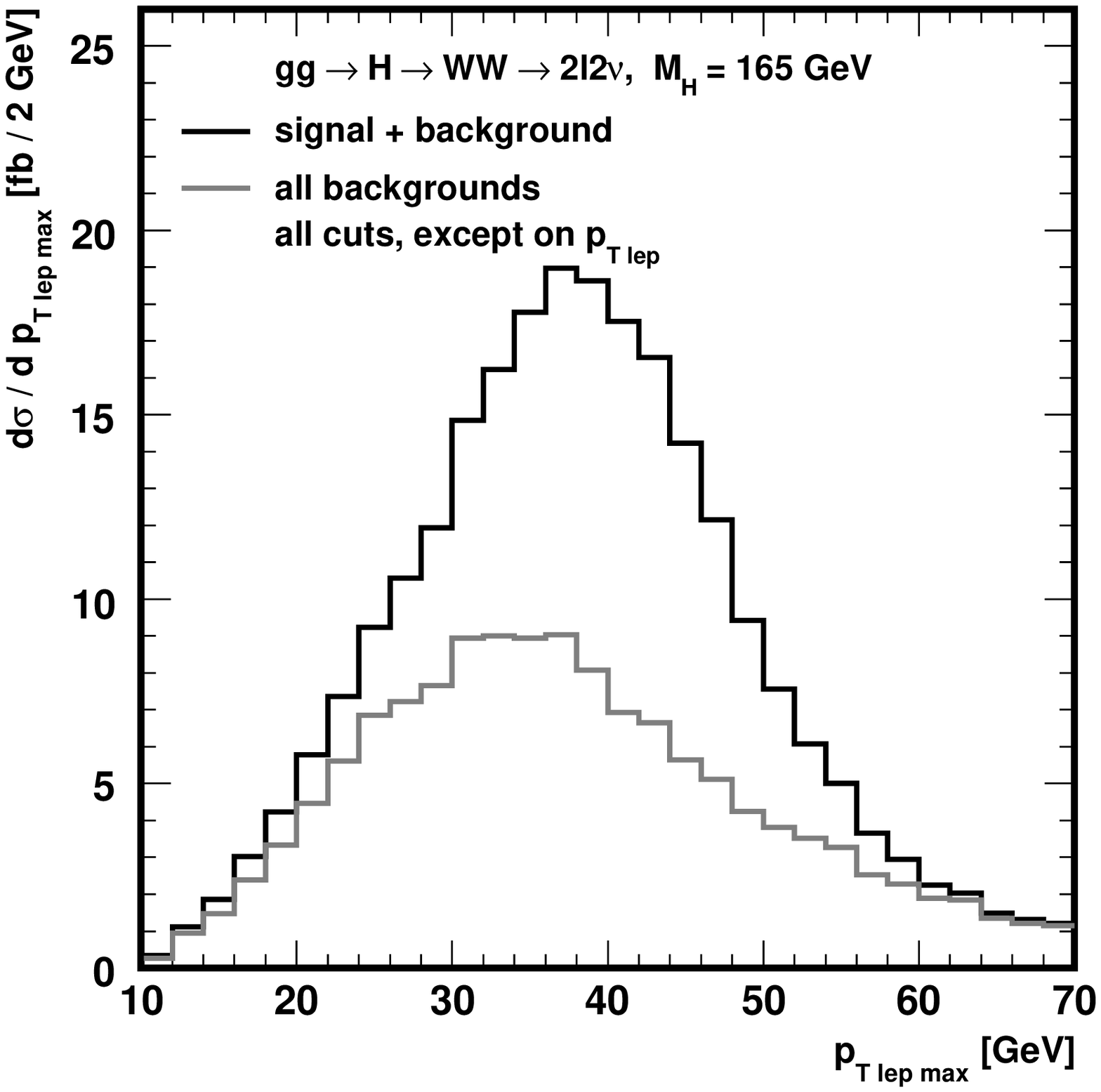}
\includegraphics*[scale=0.4]
{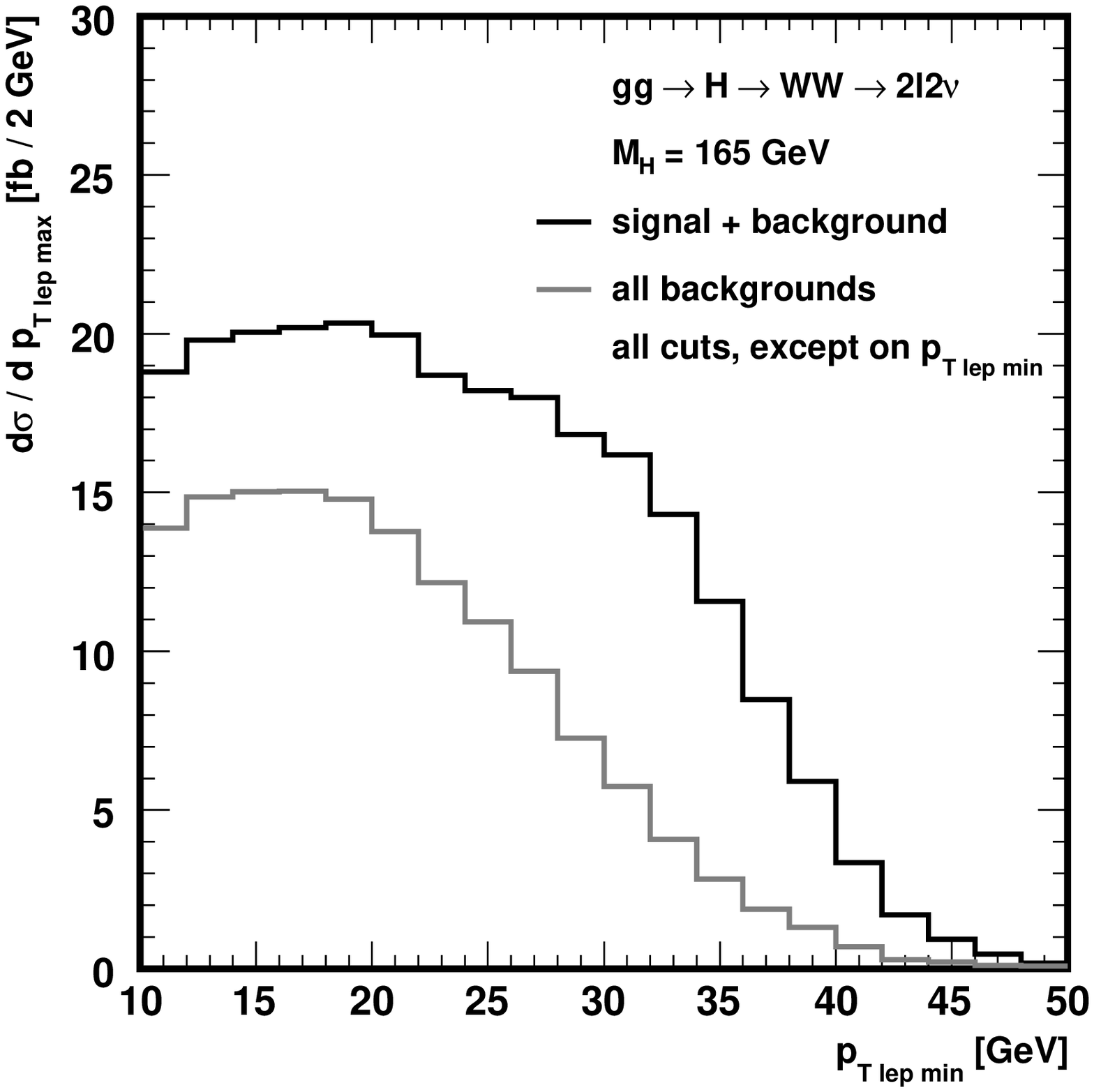}
\caption{Cross section of the Higgs signal ($\rm{M_{H}}$ = 165 GeV) and the sum of the main backgrounds, that is qqWW, ggWW, $\rm{ t\bar{t}}$ and Wtb, 
and the background alone, as a function of the \pt\ of the lepton with the maximal \pt\ (left) and the minimal \pt\ (right). All cuts are 
applied except the ones on the \pt\ of the leptons. The events are generated with PYTHIA and TOPREX and reweighted to NNLO and NLO respectively. WW
production via gluon fusion was generated using a Monte Carlo provided by N. Kauer \cite{Binoth:2005ua}, with parton shower
simulation in PYTHIA.}
\label{p4}
\end{center}
\end{figure}

\begin{figure}[h!]
\begin{center}\hspace*{-0.2cm}
\includegraphics*[scale=0.4]
{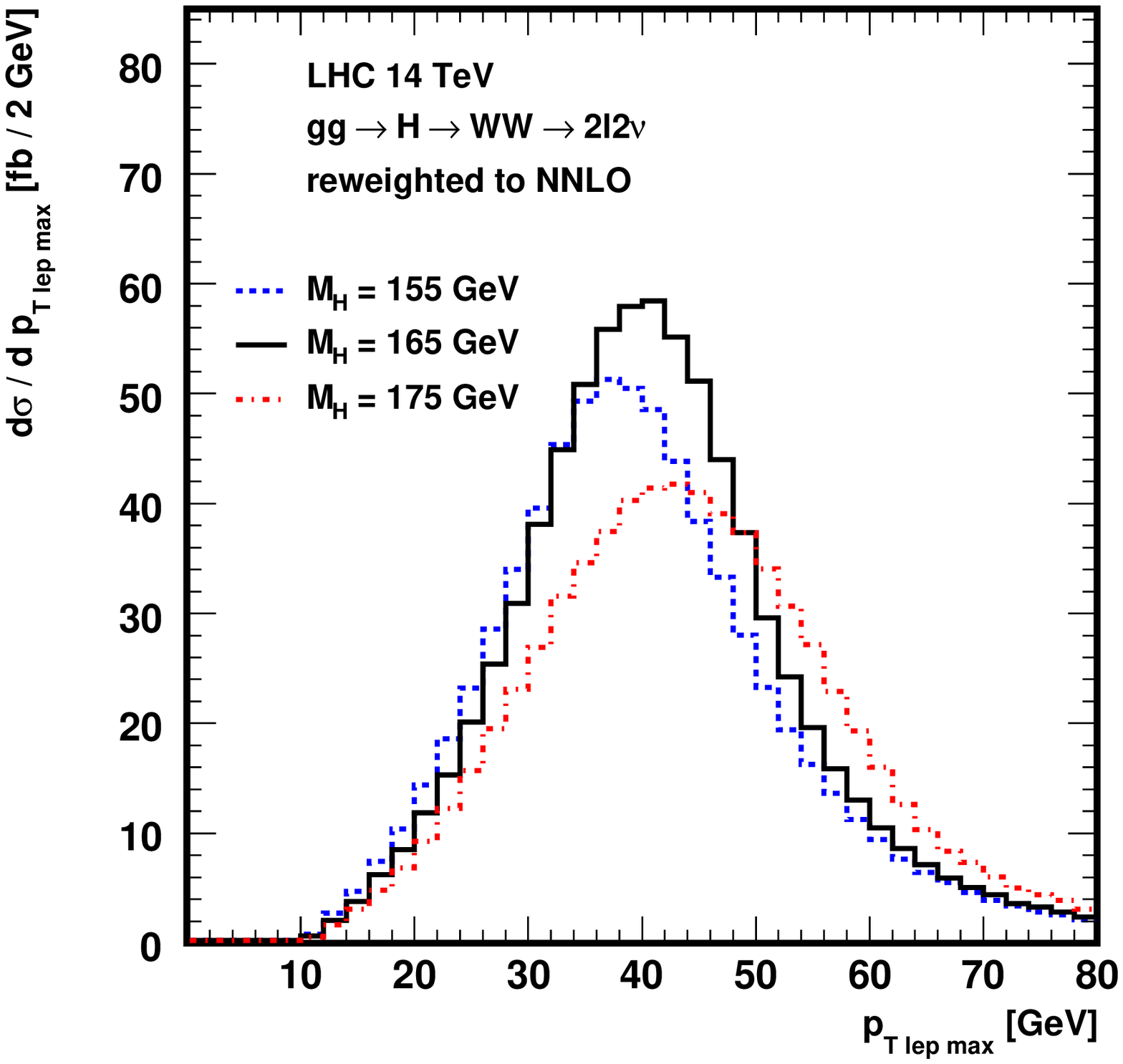}
\includegraphics*[scale=0.4]
{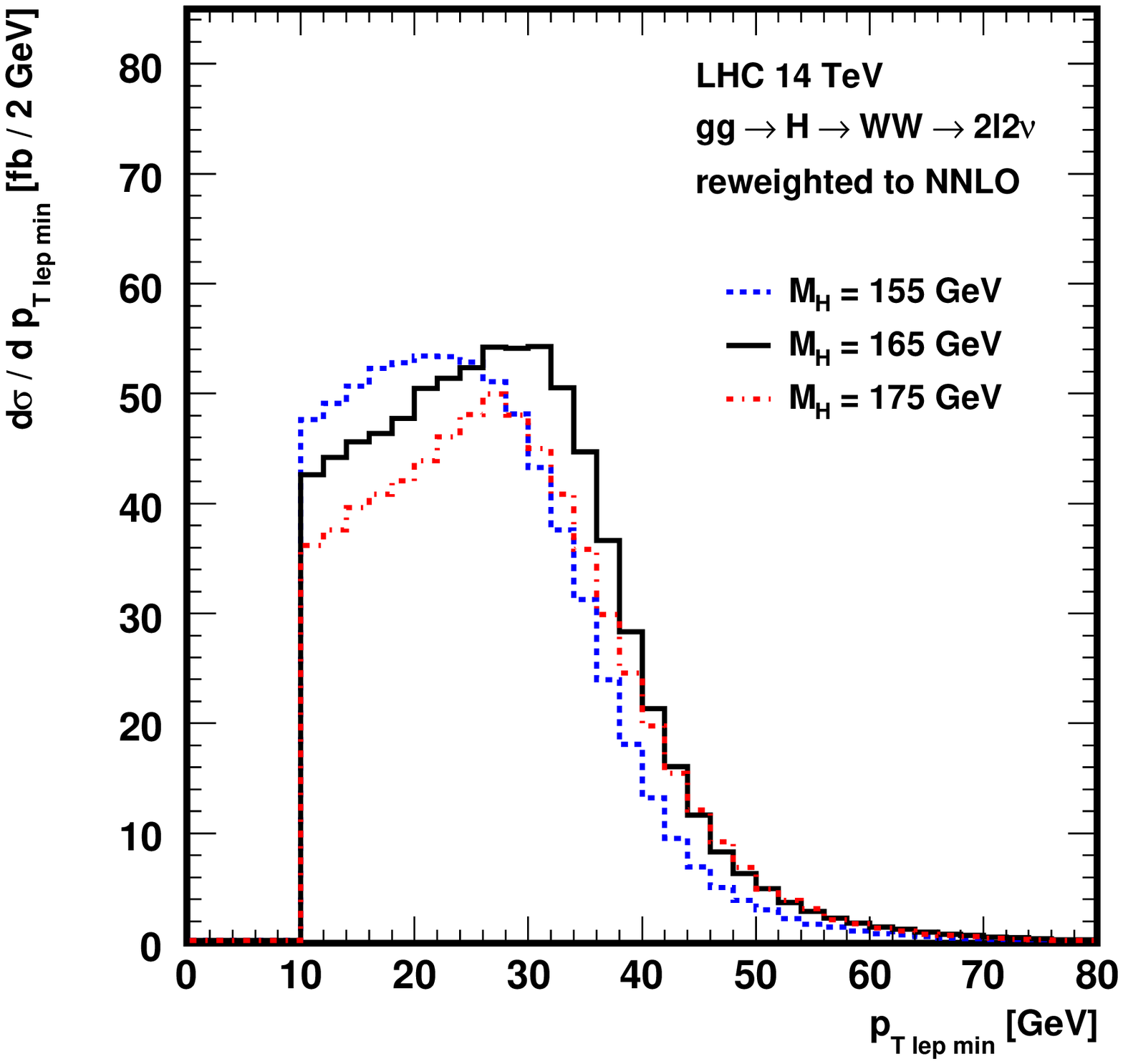}
\caption{Cross section of $\rm{gg} \fl H  \fl \rm{WW} \fl \ell\nu \ell \nu$ process for a Higgs mass of 155, 165 and 175 GeV 
as a function of the lepton \pt\ of \ptmax\ (left) and \ptmin\ (right). Only the lepton isolation cut and the minimal \pt\ of 10 GeV are applied.}
\label{p1}
\end{center}
\end{figure}

\begin{figure}[h!]
\begin{center}
\includegraphics*[scale=0.4]
{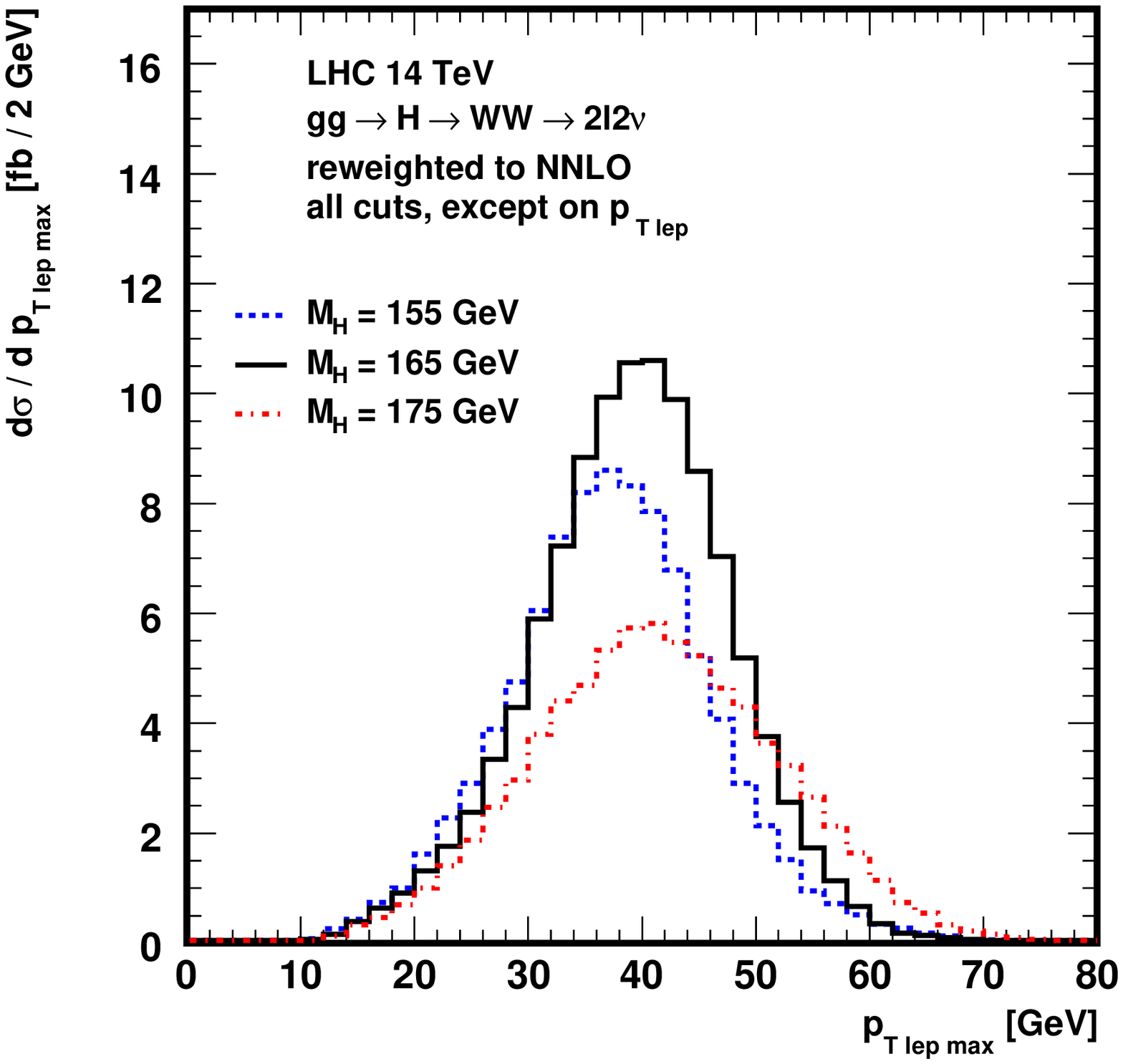}
\includegraphics*[scale=0.4]
{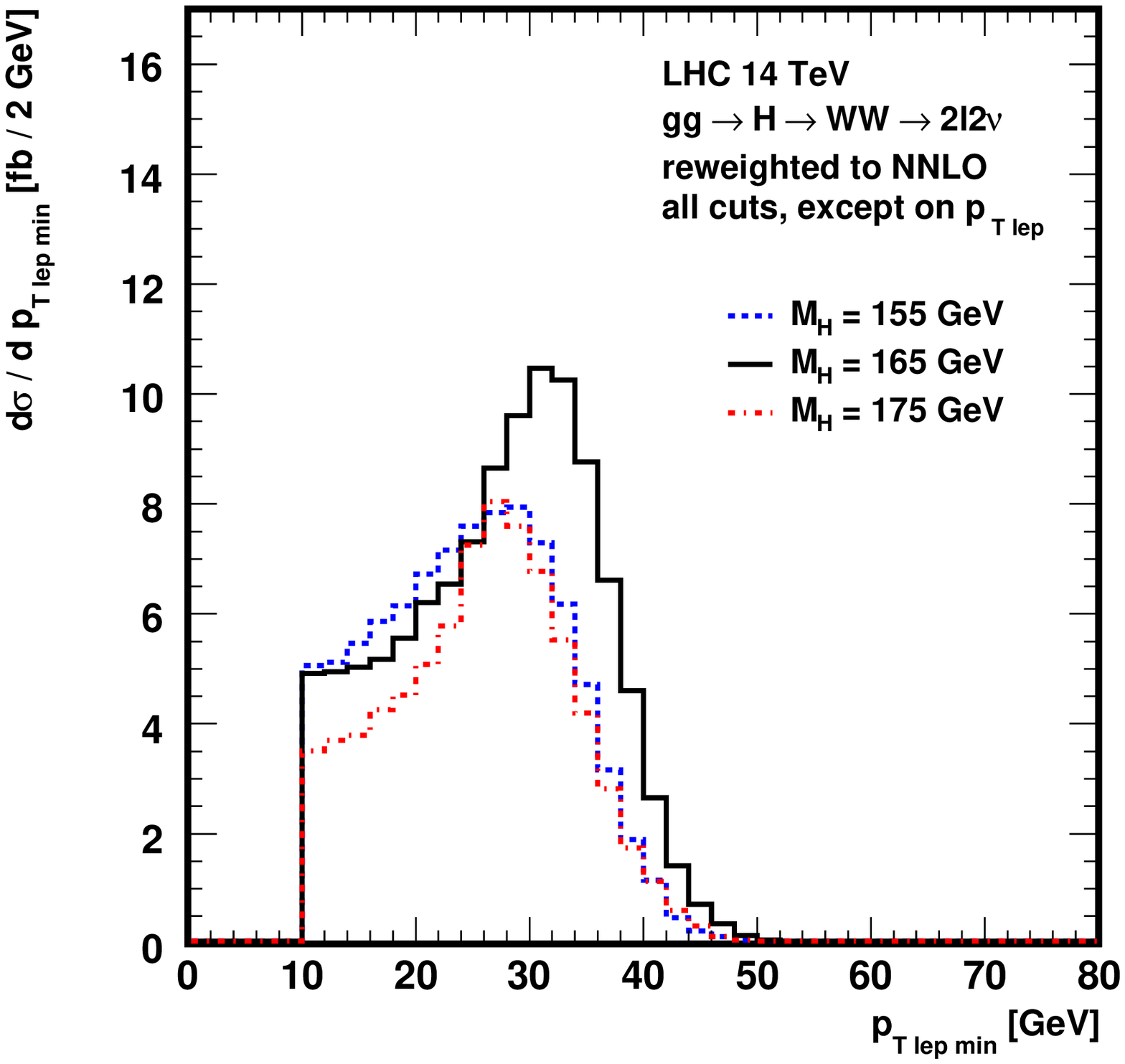}
\caption{Cross section of the Higgs signal decaying into $\rm{WW \fl \ell\nu \ell \nu}$ for a Higgs mass of 155, 165 and 175 GeV as a 
function of \ptmax\ (left) and \ptmin\ (right). All cuts, except the final ones on the \pt\ of the leptons, are applied. 
}
\label{p2}
\end{center}
\end{figure}

Since the optimal values for these cuts, especially for the cuts on the lepton \pt\ depend on the Higgs mass,
some more sophisticated tuning might be performed for other Higgs masses.
However, for the purpose of this analysis, the Higgs mass determination, we choose to use fixed selection criteria 
which provide an acceptable signal to background ratio for Higgs masses between 150 GeV and 180 GeV. 

The most relevant backgrounds are the continuum $\rm{WW}$ production \mbox{(qq $\fl$ WW $\fl$ 2$\ell$ 2$\nu$,} \mbox{gg $\fl$ WW $\fl$ 2$\ell$ 2$\nu$)} and events related to the production of $\rm{top}$ quarks \mbox{(qq $\fl$ tt $\fl$ WbWb $\fl$ 2$\ell$ 2$\nu$bb,}\mbox{ qq $\fl$ tWb $\fl$ WbWb $\fl$ 2$\ell$ 2$\nu$bb).} 
The corresponding lepton \pt\ spectra for different backgrounds 
and for a signal of $\rm{M_H}$ = 165 GeV are shown in Figure \ref{p3}a and \ref{p3}b and 
in Figure \ref{p4}a and \ref{p4}b for the sum of all backgrounds with and without the signal.\\
For the proposed cuts a signal to background ratio between 
roughly 2:1 for a Higgs mass close to 165 GeV and about 1:1 
for a mass of 150 GeV and 180 GeV is obtained.
The lepton \pt\ spectra 
for signal and background are somewhat different and depending on a more detailed 
understanding of the signal and background within future Monte Carlo simulations, some more optimized and sophisticated signal selection procedures can be done. 
For example, one sees from Figure \ref{p4}b that for \ptmin\ smaller than 20 GeV
the signal to background ratio is about 1:3. In contrast for \ptmin\  close to 30 GeV 
a signal to background ratio of about 3:1 can be obtained. 
Some further optimisation is thus certainly possible but is  
not required for the study described in the following sections. \\
Figure \ref{p1}a and \ref{p1}b show the transverse momentum spectra of the two charged leptons from the Higgs signal ($\rm{M_H}$ = 155 GeV, 165 GeV and 175 GeV), 
requiring only a minimal lepton transverse momentum of 10 GeV and   
isolation.
Figure \ref{p2}a and \ref{p2}b show the signal distributions of the lepton transverse momenta after all cuts 
except the final lepton \pt\ cuts are applied. The Higgs mass dependence, as shown in Figure \ref{p1} (before) and Figure \ref{p2} (after cuts), of the lepton \pt\ spectra remains after the selection cuts are applied. The lepton \pt\ spectra will be discussed in more detail in Section IV.

It should be clear that a convincing real data analysis needs to study in detail the 
number of accepted events for the different cuts and for signal enhanced 
and signal depleted phase space areas. 
Such an study has been performed in a recent detailed full detector 
simulation study in CMS,~\cite{Davatz:2006kb}. This study has demonstrated that 
such a model independent data driven analysis allows to constrain the $\rm{t\bar{t}}$ background with an accuracy of about 16\%. 
In a similar approach the background from $\rm{WW}$ continuum events 
have been estimated. 
A combination of these background uncertainties, weighting the 
relative errors and adding them quadratically resulted in a total systematic background error of 10\%. 
In the following we use this result from Ref.~\cite{Davatz:2006kb}
to investigate how well a potential Higgs signal cross section can be determined.

\section{Accepted Signal cross section and the Higgs mass}
\label{sec:signalsigma}

The number of signal events is the product of the theoretical signal cross section, the luminosity and the detection efficiency.  Table I summarizes the Standard Model Higgs signal cross section, the efficiency after all cuts are applied, and the number of expected events for an integrated luminosity of 10 fb$^{-1}$. One can see that the expected number of events depends strongly on the Higgs mass.

In order to use this information, the uncertainties from background, 
from the detection efficiency, the luminosity and the  
theoretical signal cross-sections have to be estimated.
\begin{table}[htb!]

\begin{center}
\begin{tabular}{|c|c|c|c|}

\hline
signal $\rm{H \fl WW} \fl \ell\nu \ell \nu$ & & & \\
\hline
$M_{H} [GeV]$ & $\sigma$ [pb] & $\epsilon$ [\%] & accepted events per 10 fb$^{-1}$\\
\hline
150&1.93&1.07&206 $\pm$9.9\%(stat.)$\pm$10.2\%(syst.)\\
\hline
155&2.08&1.45&302 $\pm$7.5\%(stat.)$\pm$7.0\%(syst.)\\
\hline
160&2.23&2.07&464 $\pm$5.6\%(stat.)$\pm$4.5\%(syst.)\\
\hline
165&2.23&2.18&486 $\pm$5.4\%(stat.)$\pm$4.3\%(syst.)\\
\hline
170&2.12&1.69&358 $\pm$6.7\%(stat.)$\pm$5.9\%(syst.)\\
\hline
175&1.96&1.28&250 $\pm$8.6\%(stat.)$\pm$8.4\%(syst.)\\
\hline
180&1.82&0.99&179 $\pm$11.0\%(stat.)$\pm$11.8\%(syst.)\\
\hline
\hline
background process &  &  &\\

\hline
qq $\fl$ tt $\fl$ WbWb $\fl$ 2$\ell$ 2$\nu$bb & 86 & 0.004&30.3$\pm$16\%\\
\hline
 qq $\fl$ tWb $\fl$ WbWb $\fl$ 2$\ell$ 2$\nu$bb& 3.4 &0.026&8.7$\pm$20\%\\
\hline
 qq $\fl$ WW $\fl$ 2$\ell$ 2$\nu$& 9.09&0.113&103 $\pm$13\%\\
\hline
gg $\fl$ WW $\fl$ 2$\ell$ 2$\nu$& 0.48 &1.473&70 $\pm$30\%\\
\hline
Combined & 99.3  & &211 $\pm$ 10\%\\
\hline
\end{tabular}\vspace{0.3cm}
\end{center}
\caption{The expected cross section for a SM Higgs at NNLO with different masses and for the 
dominant backgrounds at NLO (except ggWW which is only known at LO) are given.
The efficiencies and the number of accepted events for a luminosity of 10 fb$^{-1}$ are also given.
Both, the theoretical uncertainty for the cross section, $\Delta \sigma / \sigma$ and the   
signal efficiency uncertainty, $\Delta \epsilon / \epsilon$ are currently estimated to be about 5\%. 
The uncertainties from background subtraction systematics 
depend on the signal to background ratio and are  
is summarized in the last row.}
\label{exp}
\end{table}

It is obviously impossible to know exactly how well such measurements can be performed using 
recorded data at LHC. Nevertheless, assuming that the detectors can be operated 
as well as previous high energy hadron collider experiments, the achievable systematic uncertainties can be estimated.
The experimental uncertainties are listed first:

\begin{itemize} 
\item[1.]For the background uncertainty we use the results from a recent ``data'' driven 
CMS analysis where an accuracy of 10\% has been found~\cite{Davatz:2006kb}. 

\item[2.]For the signal efficiency it can be assumed that the efficiency for isolated high \pt\ leptons can be 
controlled from the data, using the inclusive $\rm{Z \ra ee}$ and $\rm{Z \ra \mu\mu}$
samples. Such a procedure should allow to control the charged lepton detection efficiency uncertainty 
with uncertainties of perhaps $\pm$ 1\% but certainly much smaller than 5\%. 

\item[3.]Other efficiency uncertainties come from the Monte Carlo modeling of the assumed Higgs \pt\ spectrum, 
from the rapidity dependence and from the jet activity in signal events.
We assume that these efficiencies can be measured to some extend also from  
data using various control samples like leptonic decays of inclusive W and Z events. Especially the Z events, with and without jets, allow to study in detail the underlying event with high precision, as those events have almost no background. Furthermore, the Z \pt\ spectrum can be measured from the leptons alone and with very good accuracy. A detailed analysis of the jet activity in the Z+X events, especially with a \pt\ jet close to the cut value used for the jet veto, allows to ``calibrate`` the jet veto efficiency from the data with accuracies well below 5\%. \\
The proposed signature requires only the identification of two isolated charged leptons and a veto against jets with a \pt\ above 30 GeV. As this signature is much simpler than the signal selection for $\rm{t\bar{t}}$ events, we can use a recent detailed CMS analysis of the $\rm{t \bar{t}}$ cross section measurement \cite{D'Hondt:2006ds} to estimate an upper limit for the Higgs signal efficiency uncertainty. The signal for $\rm{t\bar{t}}$ events requires at least one isolated lepton, some jets with a large invariant mass and with a possible additional requirement that one jet is tagged as a b-flavoured jet. The CMS analysis 
concluded that the $\rm{t\bar{t}}$ signal efficency can be determined with a total systematic uncertainty of 9\%, dominated 
by the b-tagging efficiency uncertainty of 7\%. The remaining uncertainties from other sources 
were estimated with an uncertainty of roughly 5-6\% . One can thus conclude that the much simpler Higgs signal signature, with two isolated high \pt\ leptons only, can be selected 
with an accuracy of $\pm$5\% or better.
\end{itemize} 

Combining these mostly experimental errors, one finds that the backgound uncertainties of 10\% ($\Delta B/B$),
varying with the particular signal to background ratio, and the efficiency uncertainty of 5\%  
match roughly the expected statistical errors already for a luminosity of 10 fb$^{-1}$. 
\def\hname{{\sf ~FEHiP}}

In order to make an interpretation of the accepted Higgs signal cross section, theoretical predictions and the luminosity uncertainties have to be estimated as well. The main contributions are the following: 

\begin{itemize} 
 
\item[1.]Today, the signal cross section is known with NNLO accuracy from perturbation theorie.
The uncertainty from unknown higher order calculations was estimated in Ref.~\cite{Moch:2005ky},
where it was shown that the calculations are converging rapidly  
and that the difference between the Higgs cross section at N$^{3}$LO and NNLO is at most about 5\%.

\item[2.]The absolut luminosity uncertainty is in general believed to be known with an accuracy 
of about 5\%. A much more accurate relative luminosity and cross section measurement, 
as proposed in \cite{Dittmar:1997md} should allow a smaller normalization uncertainty, 
reaching eventually 1\%. This approach has been used in Ref.~\cite{Martin:2002aw}, where it was pointed out that
the uncertainties from parton distribution functions, from $\alpha_{s}$
and other related systematics are very similar for the LHC cross sections of the   
SM Higgs and for the W and that the relative rates 
are already understood with an accuracy of 2-3\%.
\end{itemize} 

Thus, one finds that the theoretical cross section interpretation of a hypothetical signal is currently dominated by the 5\% uncertainties from 
future higher order calculations, if relative cross section 
measurements for the Higgs signal and for the inclusive W and Z production are used. 

We now use the expected number of signal events, as given in Table I, to 
discuss how these hypothetical signals can be used for a 
signal cross section measurement and its interpretation with respect to the Higgs mass.

The statistical error of a cross section measurement is defined as $\Delta S/ S = 1/\sqrt{S+B}$. 
As can be seen from Table I, the analysis will reach statistical uncertainties between 5-11\% for an integrated luminosity of about 10 fb$^{-1}$.
It is straightforward to use the results from Table I to estimate the statistical uncertainties for 
any other integrated luminosity. 

To summarize, the systematic uncertainties from backgrounds ($\Delta B / B$), currently estimated to 
be about 10\% for an integrated luminosity of 10 fb$^{-1}$, seem to be the dominant contribution to the systematic uncertainty of the number of events. 
For a signal to background ratio of 1:1 this would correspond to a 10\% signal uncertainty. 
For a better signal to background ratio of 2:1 the overall uncertainty would drop to 5\%.

\begin{figure}[htbp]
\begin{center}
\hspace*{-0.2cm}
\includegraphics*[scale=1.85]
{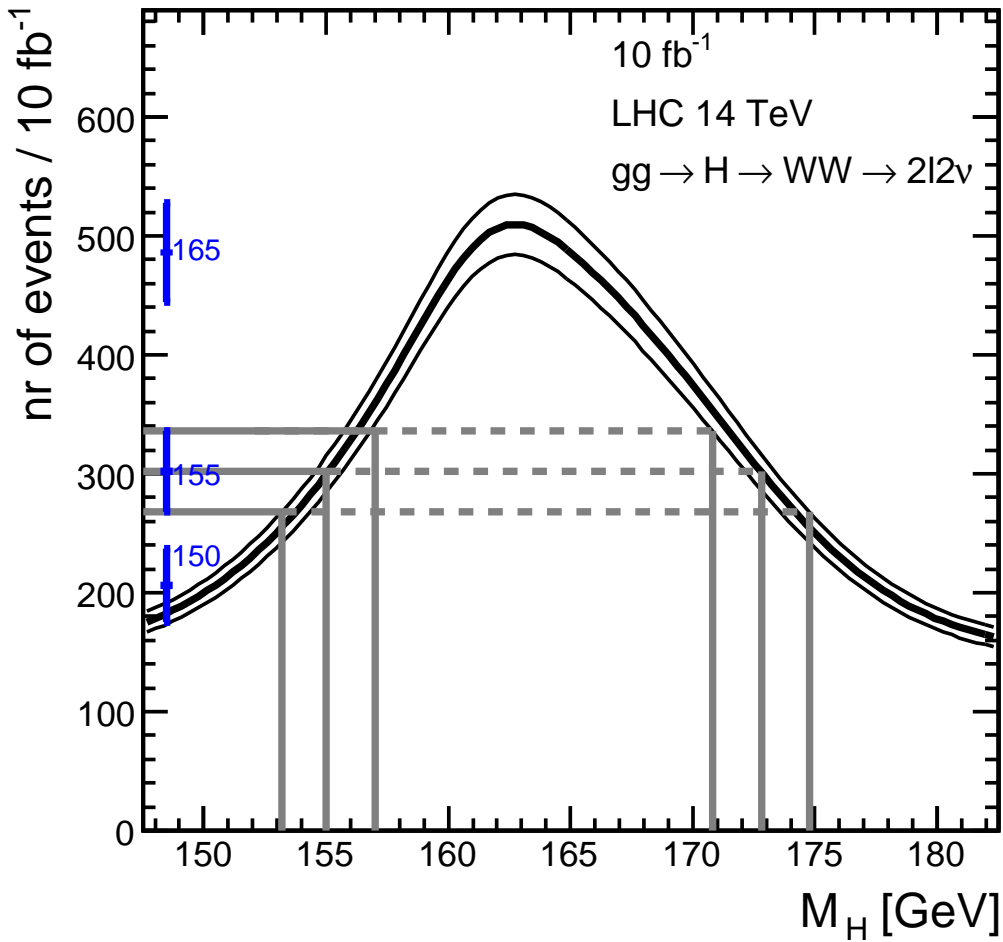}
\caption{Number of SM Higgs events events for 10 fb$^{-1}$ with all cuts applied and approximated with a fit of two Gaussian distributions
for Higgs masses between 150-180 GeV and assuming a theoretical cross section uncertainty of $\pm$ 5\%.  
Three hypothetical experimental numbers of signal events, including statistical and systematic error, are also shown. For the case 
of a hypothetical result corresponding to a Higgs mass of 155 GeV,  
the possible graphical interpretation in terms of the SM Higgs mass is also indicated.}
\label{p5}
\end{center}
\end{figure}

Larger data sets, combined with a well understood 
detector and more accurate MC generators might result in some reduction of this background uncertainty. 
However, such future improvements can easily be included into the strategy and the results obtained with the current assumptions.
The uncertainty from the detection efficiency, as discussed above, can be expected to be about 5\%.
In addition one must take the uncertainty from the theoretical prediction into account. It is currently assumed that, 
using relative measurement and optimal cross section ratios, an accuracy of 5\% has been reached already.
Thus, combining all these errors an entire systematic uncertainty of 10-15\% for the cross section measurement seems to be realistic. The systematic uncertainty starts to be the dominant error once a luminosity of about 10-20 fb$^{-1}$ is reached.
Figure \ref{p5} shows the number of hypothetical signal events within the Standard Model 
and for an integrated luminosity of 10 fb$^{-1}$.

In order to demonstrate how such a ``result'' can be interpreted in terms of the Higgs mass, the theoretical expectation for accepted Higgs signal events was determined for different Higgs masses such that an analytic function could be used to approximate the prediction for all Higgs mass values. The accepted events were obtained with PYTHIA reweighted to the HqT program~\cite{Bozzi:2003jy,Bozzi:2005wk}. A fit with two Gaussian functions was used to obtain the curve. 
The experimental and  theoretical uncertainties will be split in the following way.
The band indicates the $\pm$ 5\% theoretical cross section uncertainty.
As can be seen from Figure \ref{p5}, a broad maximum of signal events is predicted for a mass between 160 and 168 GeV. For lower and higher masses, the expected rate of signal events decreases steeply.
On the left side of the Figure, the estimated number of accepted events for three different Higgs masses, as listed in Table I, are shown.
The corresponding number of events are  206 $\pm$ 29 for a Higgs mass of 150 GeV,
302 $\pm$ 35 events for a Higgs mass of 155 GeV and 486 $\pm$ 34 for 165 GeV.  
The statistical and the estimated experimental systematic errors 
have been added in quadrature.

Taking first the expected number of 302 $\pm$ 35 events 
the graphical interpretation leads directly to a possible SM Higgs mass of either the correct mass of  
155 $\pm$ 2 GeV or to a second solution of 173 $\pm$ 2 GeV. A similar accuracy and 
twofold ambiguity would be obtained for other hypothetical number of events outside the peak region.

For signal numbers between 460 to 480 events, the uncertainty being about $\pm$ 35 events, 
one would interpret this result with a SM Higgs somewhere between 160 and 168 GeV with nearly equal probability.

Assuming a nearly flat probability distribution for all masses within this interval, 
one could estimate that the most likely mass value would be 164 GeV with the corresponding RMS error of 2.3 GeV, 
as defined from a box-like distribution with a width of 8 GeV \mbox{(RMS = width/$\sqrt{12}$).}

\section{Higgs mass and lepton \pt\ spectra}
\label{sec:signalcuts}

In the previous section we have shown how a potential Higgs signal cross section 
measurement in the mass range between 150-180 GeV can be used for a first  
estimate of the Higgs mass. If one finds another observable which is suited to distinguish between the now ambigious mass predictions in the tails of the mass curve, an accuracy of about $\pm$ 2 to 2.5 GeV can be found, assuming that the Higgs has Standard-Model-like properties. 

Now we will try to analyse correlations between the Higgs mass, the Higgs transverse momentum 
spectrum and the observable lepton \pt\ spectra in order to find such an observable.
As has been shown in Figures \ref{p1} and \ref{p2}, the \pt\ spectrum of both leptons, before and after the selection cuts, depends on the Higgs mass.

\begin{figure}[htb!]
\begin{center}
\hspace*{-0.2cm}
\includegraphics*[scale=0.4]
{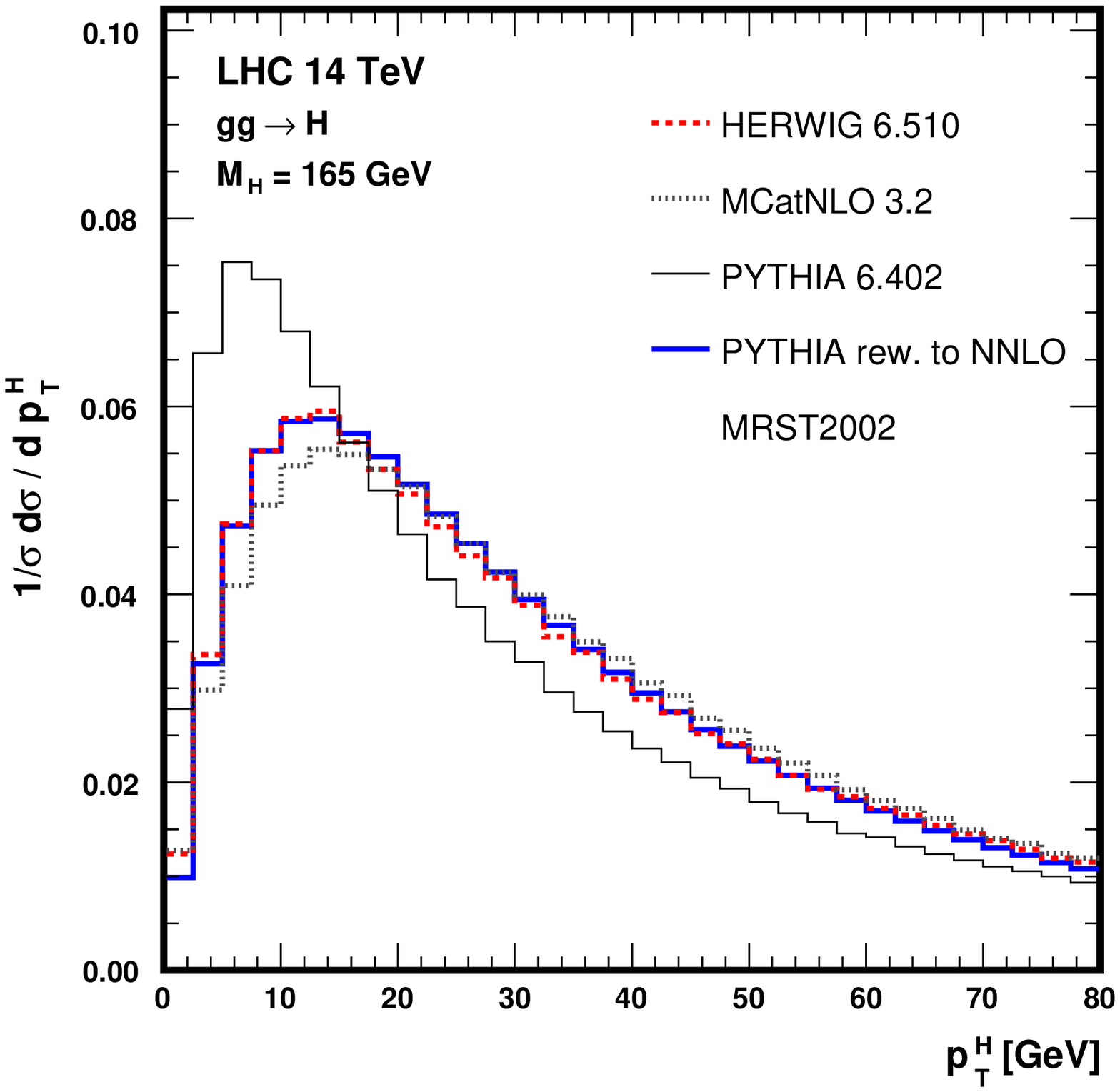}
\includegraphics*[scale=0.4]
{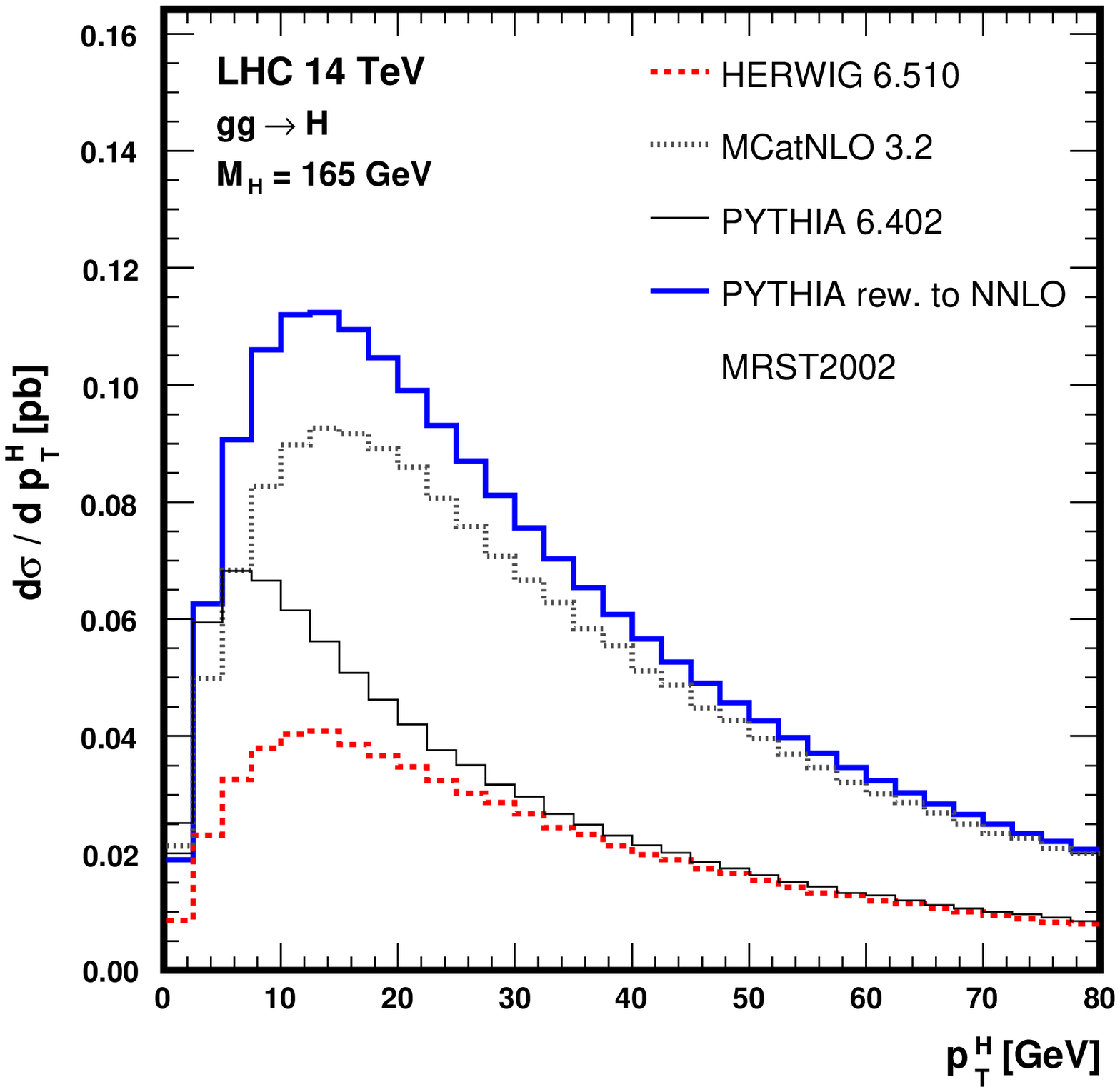}
\caption{The \pt\ Higgs spectrum for a Higgs mass of 165 GeV and four different Monte Carlo programs: HERWIG, 
MC@NLO, PYTHIA and PYTHIA reweighted to NNLO. Normalized (left) and absolute (right) scale. No cuts are applied. For PYTHIA, the q-ordered showering is used. More \pt\ Higgs spectra, including the predictions from $\rm k_T$-ordered shower in PYTHIA and results with CASCADE \cite{Jung:2001hx}, can be found in \cite{Alekhin:2005dx} p.246-250 and \cite{Buttar:2006zd} p.127-131.}
\label{p6}
\end{center}
\end{figure}

Figure \ref{p6}a shows the Higgs cross section as a function of the Higgs \pt\ for different Monte Carlo predictions ($\rm{M_H}$ = 165 GeV). Compared are PYTHIA, HERWIG\cite{Corcella:2000bw}, 
MC@NLO \cite{Frixione:2003ei}, and HqT. Figure \ref{p6}b shows the normalized cross section to compare the shapes of the different calculations. 

For Higgs transverse momenta smaller than 10-15 GeV, large unknowns prevent currently an accurate calculation of the Higgs \pt\ spectrum.

The used search strategy requires various selection criteria, especially the jet veto (to remove top-like events), 
which essentially remove all events where the Higgs has a transverse momentum larger than the value for the jet veto cut (\ptJ\ ). In this analysis, \ptJ\ = 30 GeV was used. Therefore, the low \pt\ Higgs region is especially important for this signature. 

\begin{figure}[htb!]
\begin{center}
\hspace*{-0.2cm}
\includegraphics*[scale=1.02]
{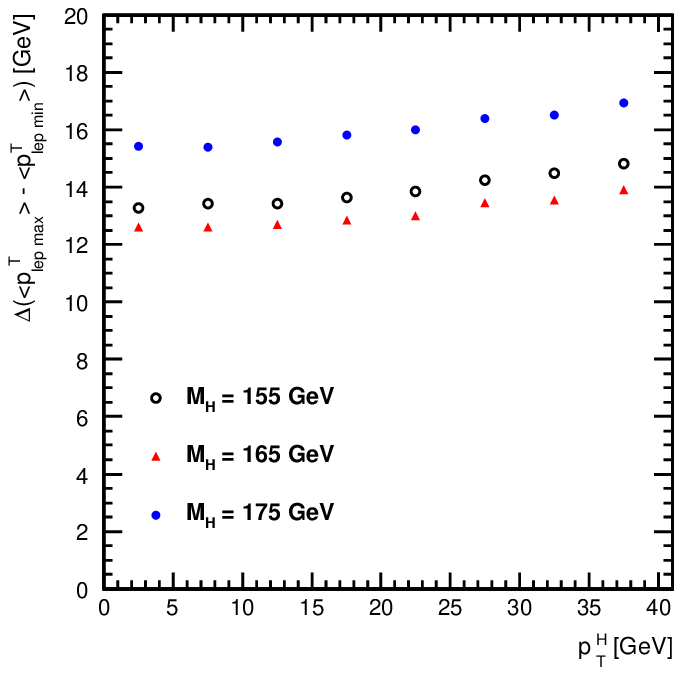}
\hspace*{0.7cm}\includegraphics*[scale=1.02]
{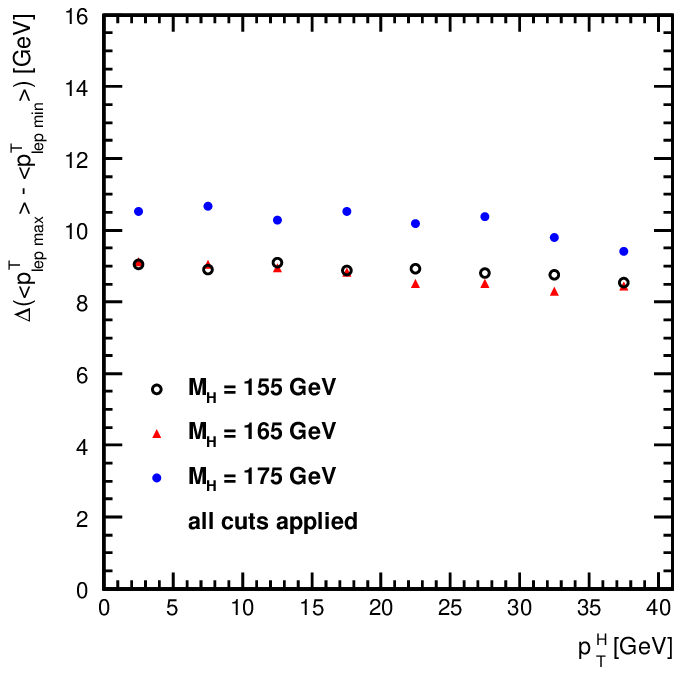}

\caption{Average $\Delta \pt\ $  between the leptons as a function of \pt\ Higgs. (Left) No cuts are applied 
except two isolated leptons required (right) all cuts are applied.}
\label{p7}
\end{center}
\end{figure}

\begin{figure}[htb]
\begin{center}
\hspace*{-0.2cm}
\includegraphics*[scale=0.4]
{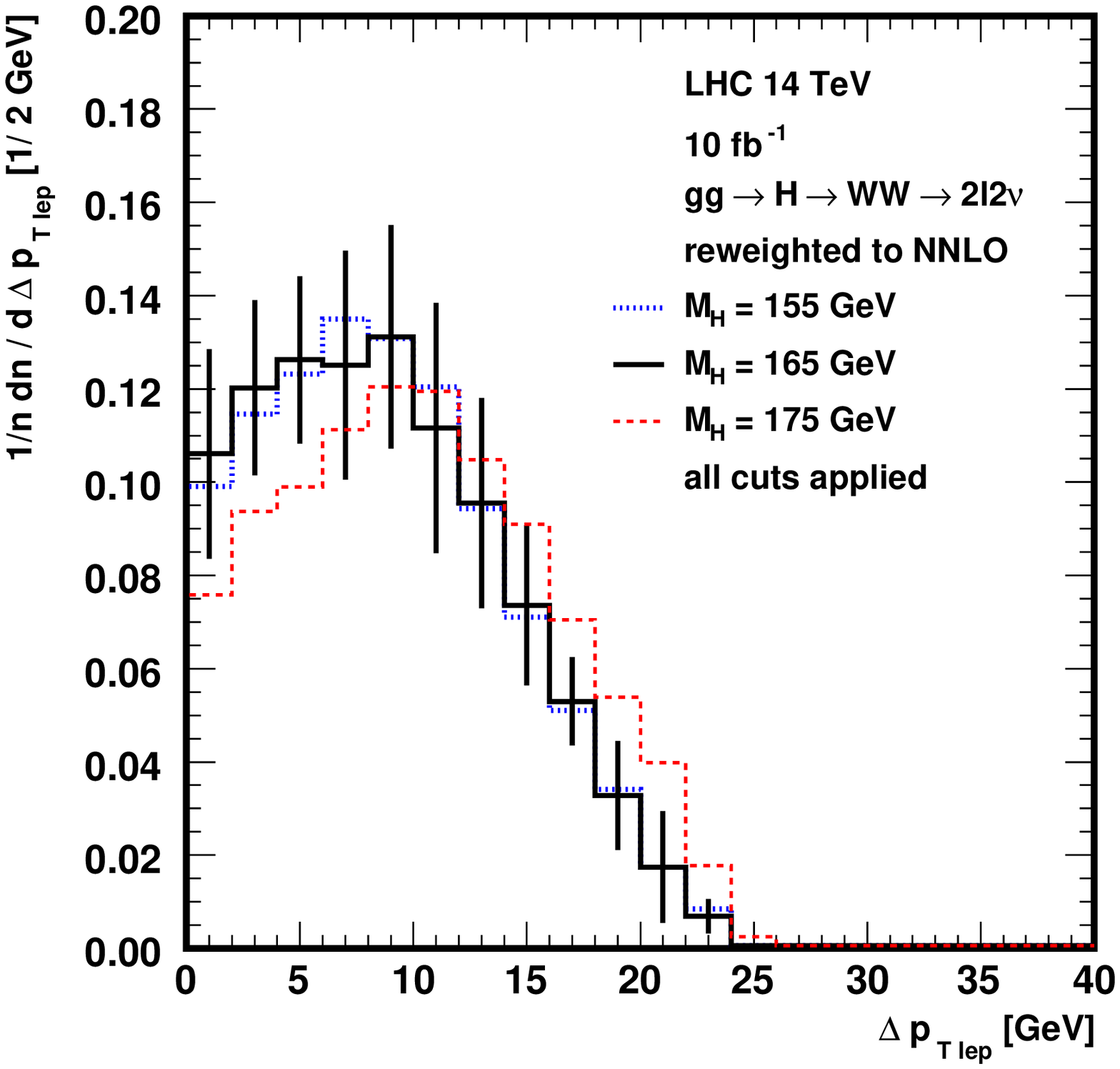}
\includegraphics*[scale=0.4]
{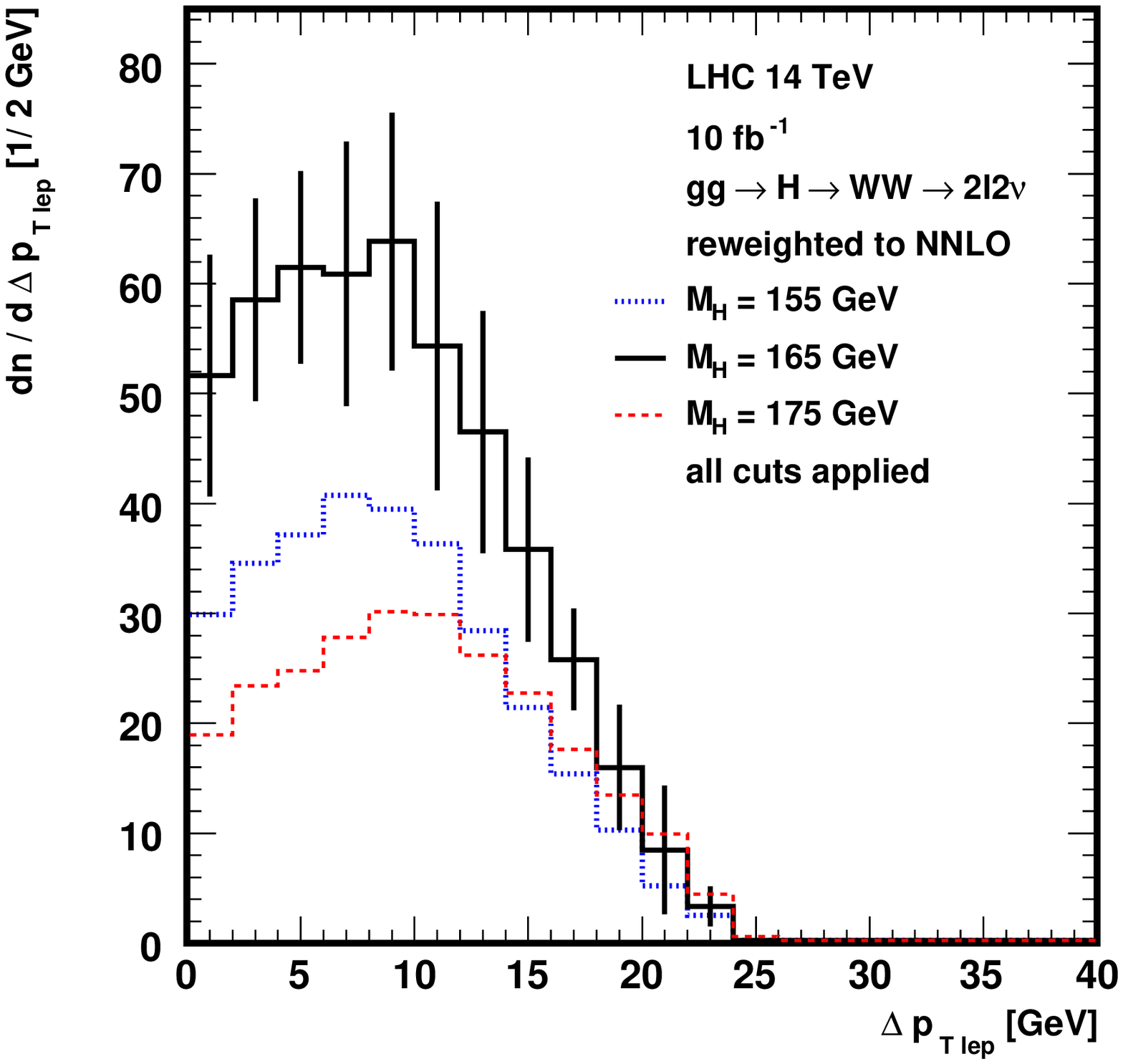}
\caption{Difference between the \pt\ distributions of the leptons for different Higgs masses. The error for $\rm{M_H}$ 165 GeV is given for a luminosity of 10 fb$^{ -1}$. (Left) Normalized (right) Number of events for 10 fb$^{ -1}$. All cuts are applied.}
\label{p8}
\end{center}
\end{figure}

\begin{figure}[htb]
\begin{center}
\hspace*{-0.3cm}
\includegraphics*[scale=1.02]
{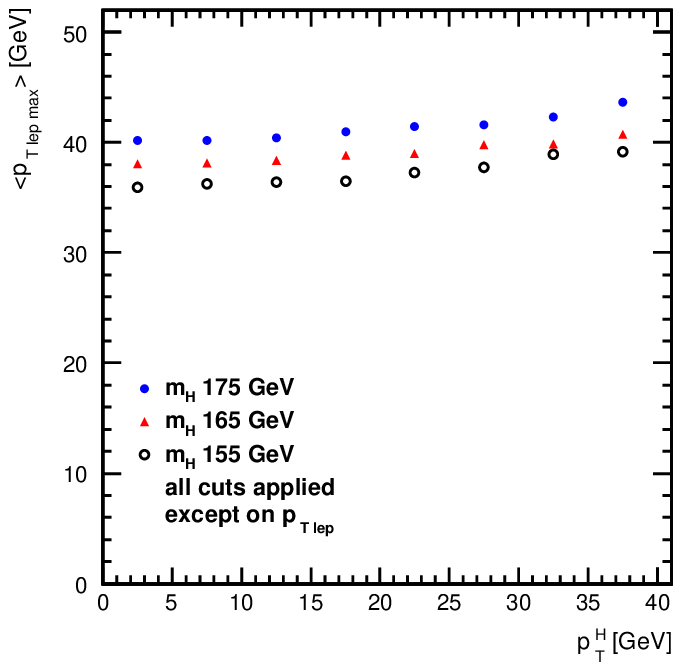}
\hspace*{0.5cm}\includegraphics*[scale=1.02]
{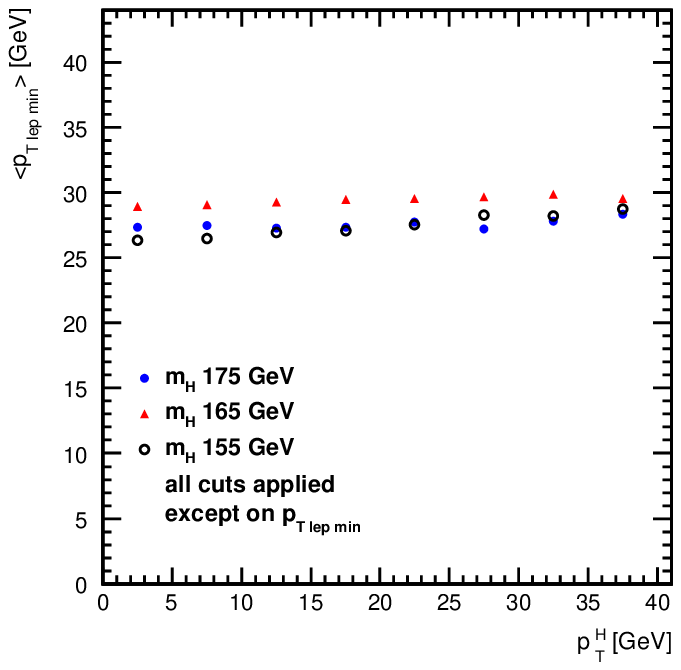}
\caption{Average transverse momenta of the leptons as a function of \pt\ Higgs, all cuts applied, except the ones on \pt\ of the leptons (left) and except the ones of \ptmin\ (right).}
\label{p9}
\end{center}
\end{figure}

\begin{figure}[htb]
\begin{center}
\hspace*{-0.2cm}
\includegraphics*[scale=0.4]
{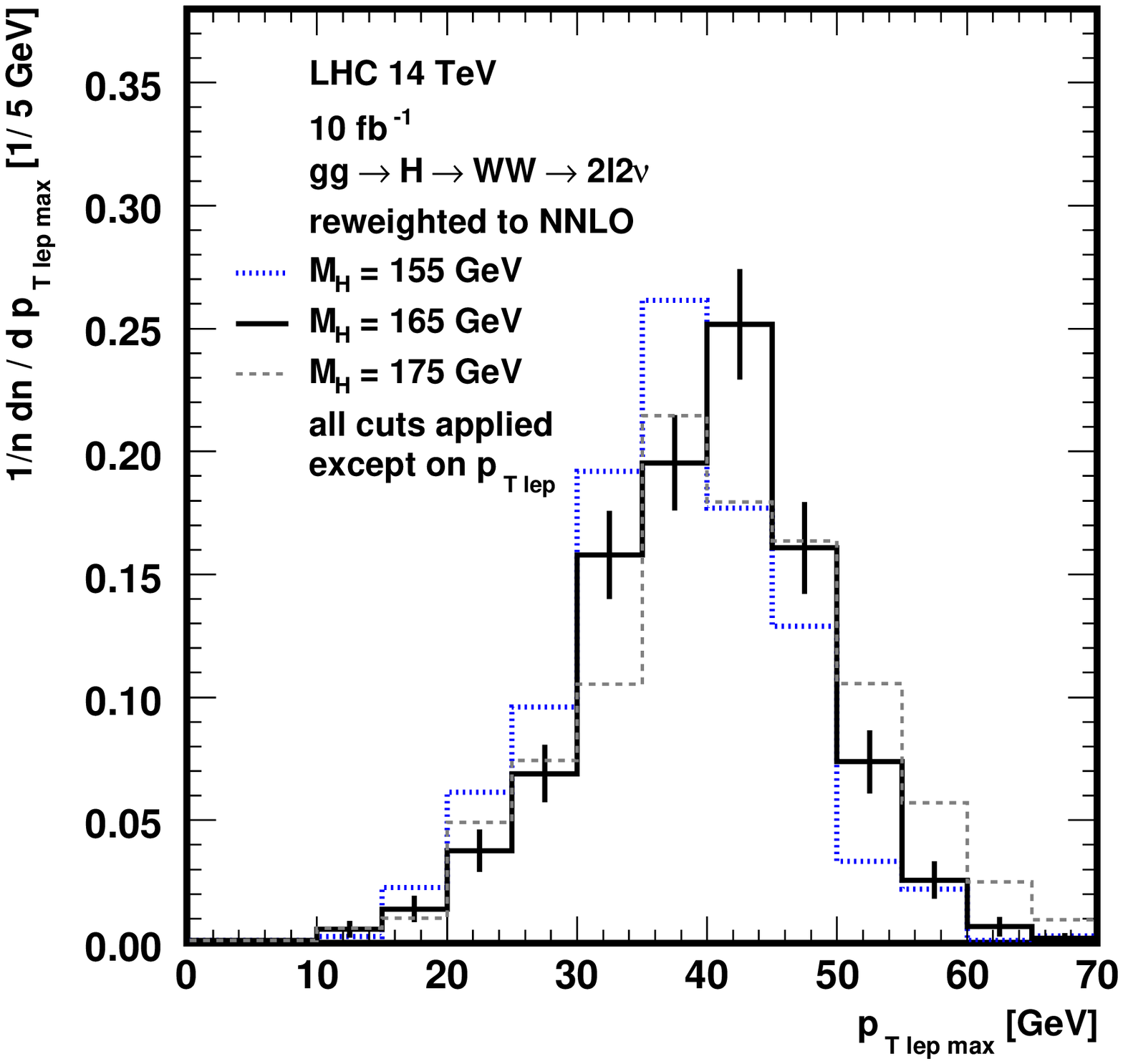}
\includegraphics*[scale=0.4]
{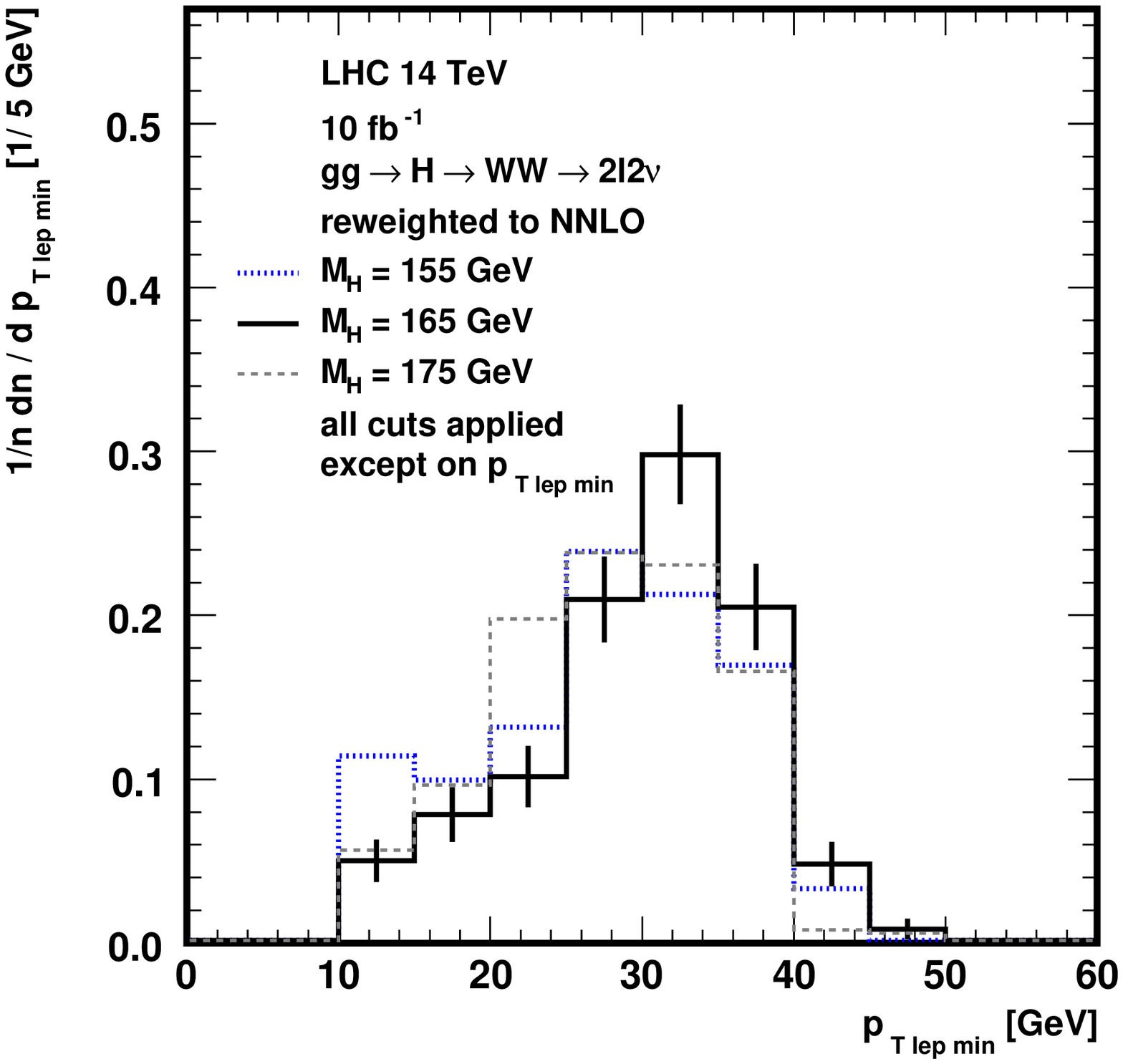}
\caption{Normalized \pt\ distribution of the leptons for different Higgs masses and for a luminosity of 10 fb$^{ -1}$. (Left) All cuts are applied except the ones on \pt\  of the leptons (right) all cuts are applied except the one on \ptmin.}
\label{p10}
\end{center}
\end{figure}

\begin{figure}[htb]
\begin{center}
\hspace*{-0.2cm}
\includegraphics*[scale=0.4]
{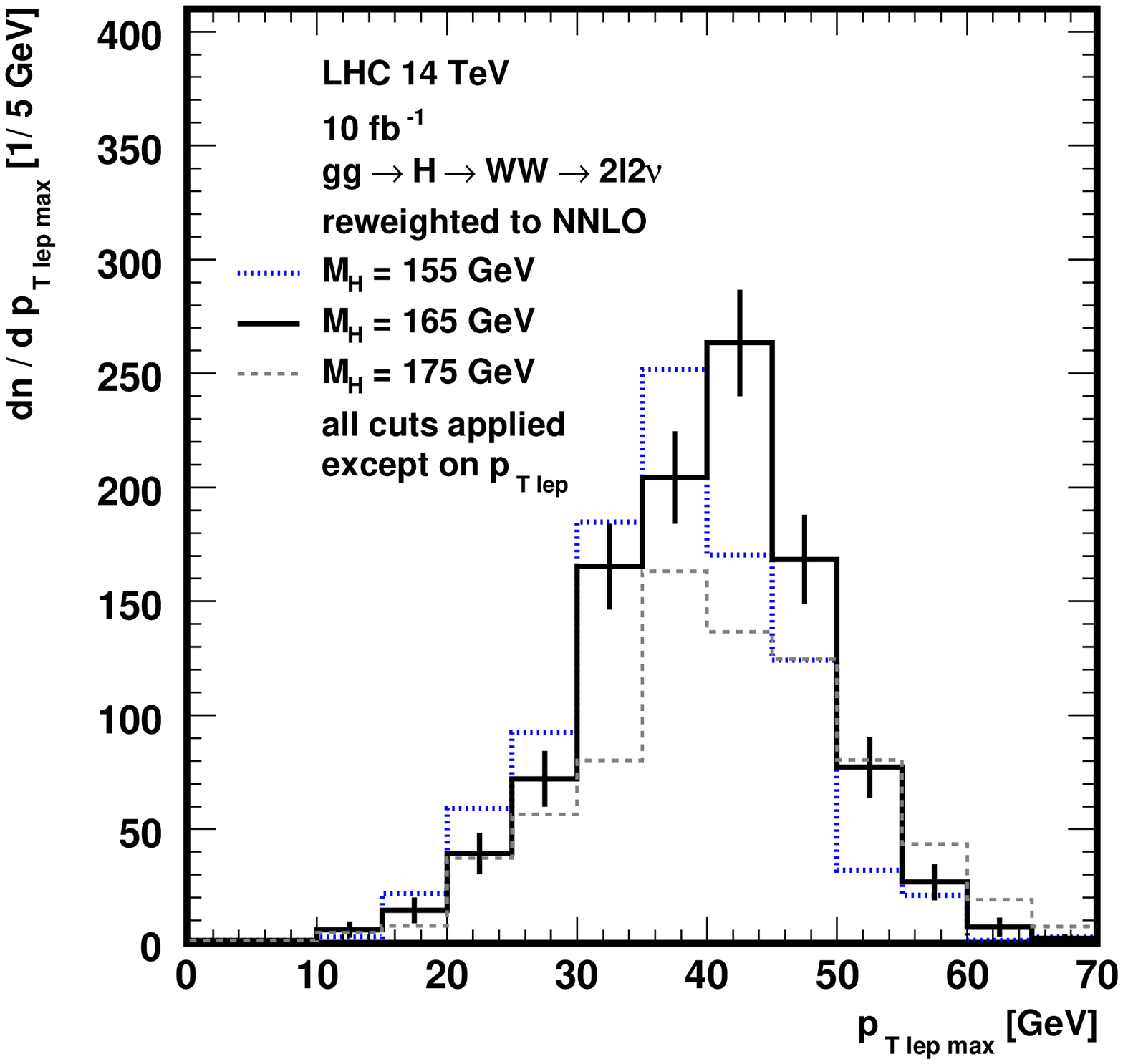}
\includegraphics*[scale=0.4]
{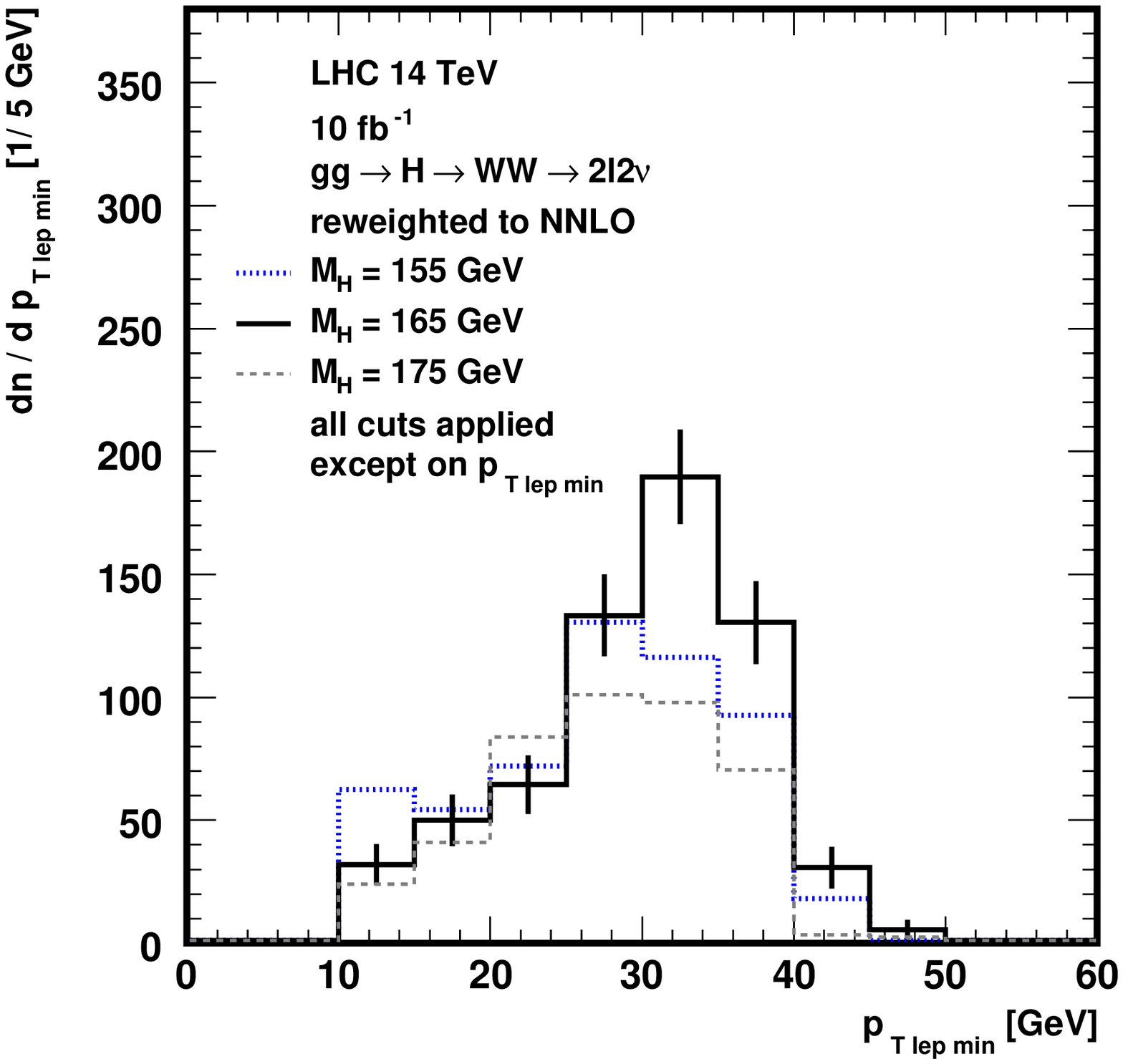}
\caption{\pt\ distribution of the leptons for different Higgs masses and for a luminosity of 10 fb$^{ -1}$. (Left) All cuts are applied except the ones on \pt\  of the leptons (right) all cuts are applied except the one on \ptmin.}
\label{p11}
\end{center}
\end{figure}

Figure \ref{p6}a and \ref{p6}b demonstrate that the Higgs \pt\ spectrum can currently not be calculated 
from first principles. In addition, the Higgs \pt\ spectrum can not be measured directly in this channel. One should therefore try to find an observable which is rather independent from the specific Higgs \pt\ spectrum. We will investigate in the following if the Higgs mass sensitivity of the lepton \pt\ spectra
is large enough compared to the potential correlations with the not well known Higgs \pt\ spectrum. 

The studied experimental observables are the lepton \pt\ spectra (\ptmax\ and  \ptmin\ ) and $\Delta$\pt\, defined on an event by event basis (\ptmax\ - \ptmin\ ). 
 
Figure \ref{p7} shows the mean value of $\Delta$\pt\ without (\ref{p7}a) and with (\ref{p7}b) cuts as a function of the Higgs \pt\ and 
for a Higgs mass of 155 GeV, 165 GeV and 175 GeV.
It is interesting to see that this observable, before and after cuts, is essentially independent from the Higgs \pt\ spectrum. Consequently, the unkown details of the Higgs \pt\ spectrum are not relevant for the interpretation of this observable. We will investigate now how well the predictions for a Higgs mass of 155 GeV and 175 GeV can be distinguished. Figure \ref{p7}b shows that the expected mean values between a Higgs mass of 155 GeV and 175 GeV. With all cuts applied 
the \pt\ difference of the two charged leptons is on average 
about 10 GeV for a Higgs mass of 175 GeV compared to about 9 GeV for a mass of 165 GeV and lower.
This difference of about 1 GeV should be compared to the error for the mean value, expected to be 
roughly  RMS/$\sqrt{N}$, or about 5.3 GeV/$\sqrt{302}$ = 0.3 GeV for an integrated luminosity of 10 fb$^{-1}$ and a mass of 155 GeV. Consequently, the distribution allows to distinguish a mass of 155 GeV and 175 GeV with at least 3 sigma.

Next, the shape of the entire $\Delta$\pt\ distribution, shown in Figure \ref{p8}a (normalized) show the predicted shape and \ref{p8}b in addition the differential distribution expected 
from the Standard Model using the NNLO prediction (with reweighting) from the HqT program for 10 fb$^{-1}$ and with all cuts applied.

Using only the shapes, shown in \ref{p8}a, the statistics from a measurement with a luminosity of 10 fb$^{-1}$ should be sufficient 
to distinguish a Higgs mass of 155 GeV from one at 175 GeV independent from any cross section assumptions. 
In a more detailed analysis and once a signal has been observed, a detailed shape analysis 
would accurately count the number of signal events per $\Delta$\pt\ bin. For example one could measure the fraction of signal events 
with $\Delta$\pt\ larger than 15 GeV in Figure \ref{p8}a, resulting in a difference of about 2 sigma between 
175 GeV and the two lower mass hypotheses 165 GeV and 155 GeV. Such an analysis would certainly also benefit from much larger luminosities. 
Similar and if in addition the SM cross section prediction is assumed, as shown in Figure \ref{p8}b, even a mass between 
165 GeV and 155 GeV can be separated easily.

Finally, one can study how the lepton \ptmax\ and \ptmin\ distributions are correlated with the Higgs mass and the Higgs \pt\ spectrum.  
Figure \ref{p9}a and \ref{p9}b show (1) that the average values for \ptmax\ and \ptmin\ are essentially independent from details of the Higgs \pt\ spectrum 
and (2) that some correlation with the Higgs mass exists.  
Figure \ref{p10}a and \ref{p10}b show the shape of the \ptmax\ and \ptmin\ distributions for accepted events 
with all cuts except the final lepton \pt\ cuts for \ptmax\ distribution and all cuts except the cuts on \ptmin\ in the \ptmin\ distribution are applied. The corresponding distributions with the absolut cross section are shown in Figure \ref{p11}a and \ref{p11}b.  
These distribution demonstrate clearly that a more detailed analysis 
of the \pt\ spectrum, especially when much larger luminosities become available, can increase 
the potential mass range for this signature perhaps to Higgs masses well below 150 GeV 
and should improve the Higgs mass measurement considerable. 
For example, as one can see in Figure \ref{p11}a, about 21\% of the excess events and a Higgs mass of 175 have a 
\ptmax\ larger than 50 GeV. This should be compared with about 11\% for a mass of 165 GeV and
6\% for a mass of 155 GeV.  
Once an excess of Higgs-signal-like events has been demonstrated  
one would then analyze the distributions of \ptmax\ and \ptmin\ 
in much more detail. For example, one could 
measure the excess of signal events for \ptmax\ larger than 
50 GeV and/or \ptmin\ between 10 GeV and 20 GeV and compare this to 
the mass dependence in the most accurate Monte Carlo simulations.  

Figures \ref{p7} and \ref{p9} demonstrate that the Higgs mass and the average lepton transverse momenta are correlated, while essentially no correlation between the transverse Higgs momentum and the average lepton \pt\ can be seen.

This on the first view surprising result can be qualitatively understood as follows. The transverse momentum spectra of the leptons depend essentially only on the mass of the intermediate W, their \pt\ , and the spin correlations between the two W's.

For a Higgs mass  slightly above 160 GeV, essentially both W bosons are on shell. For lower Higgs masses one finds that it is more likely that the available decay energy will rather go into the W mass than into its momentum. For larger Higgs masses, the energy of the W boson and thus its average transverse momentum  will increase rapidly. For example in the rest frame of the Higgs with a mass of 180 GeV, decaying into two on shell W's, one expects already an average W momentum of about 40 GeV. 

Depending on the spin orientation, the charged leptons will either be emitted in the direction of the W momentum or against it in the rest frame of the Higgs. In case the lepton is emitted in the direction of the W, its momentum will also be increased. In contrast, if its emitted opposite the W momentum, as forced by the spin correlations in the W rest frame, its momentum in the lab frame will be reduced. 
As long as the mass of the W is much larger than its \pt\ , the two resulting leptons will have a small opening angle in the lab frame.

For very large Higgs masses, this condition  is not fulfilled any more, and the small opening angle between the two charged leptons, which is the signal signature, will disappear.\\
What happens now when the Higgs boson, with a mass close to 2 $\rm{M_W}$, itself has a small (compared to $\rm{M_W}$) transverse momentum?
On average and very qualitatively, this will add some transverse momentum to the W emitted in the direction
of the momentum of the Higgs boson and reduce the transverse momentum of the other W boson.
Consequently, one charged lepton gets a slightly larger \pt\ and the other one a slightly smaller \pt\  .
As a result one would expect that the width of the lepton \pt\ spectrum gets broader
while their mean values and especially their difference remain essentially unchanged as long as the
selection criteria do not cut too strongly in the low momentum tail.

\section{Combined Higgs mass measurement}
\label{sec:combination}

In section III the correlation between the number of expected signal events 
with the Higgs mass has been used to estimate the Higgs mass.
As a result a mass measurement with an error between 2.0 GeV and 2.5 GeV has been obtained 
assuming the validity of the Standard Model cross section calculation with a theoretical accuracy of 5\%.
However, the cross section interpretation for masses below 160 GeV or above 168 GeV 
leads to two possible Higgs masses. In section IV correlations between the lepton transverse momenta and the Higgs mass have been 
studied. In particular we have demonstrated that the lepton \pt\ spectra and their difference 
originating from a Higgs with a mass of 155 GeV and 175 GeV differ, in a rather model independent way, by more than 3 standard deviations.
Thus, once the number of signal events is not consistent with the mass intervall between 160-168 GeV, 
the cross section ambiguities for the mass interpretation can be resolved by a detailed analysis of the lepton \pt\ spectra.   
A combination of the accepted number of events (or accepted cross section) 
with the analysis of the lepton \pt\ spectra will allow to identify which of the mass hypotheses is valid. For example 
if about 300 signal events would be selected for a luminosity of 10 fb$^{-1}$, either a mass of 155 $\pm$ 2 GeV or 
173 $\pm$ 2 GeV would be possible. The analysis of the lepton \pt\ spectra indentifies the correct solution.
In a similar approach and once a signal has been identified, one would test if the mass interpretation from the Standard Model cross section predicts also a good agreement with the observed lepton \pt\ spectrum. If this is not the case, and with larger statistics, the lepton \pt\ spectra can be used to demonstrate that the observed signal deviates from the SM.

\section{Summary} 
\label{sec:conclusions}

Assuming that the LHC experiments will discover a Higgs-like signal in the 
channel $\rm{gg \to H \to WW \to \ell \nu \ell \nu}$, different experimental observables
have been analysed in order to establish how well the Higgs mass can be measured 
with this channel. 
Using the now well established selection procedure, the 
observed event rate can be used to determine the Higgs mass, assuming Standard Model couplings. 
In addition it is shown that the observable lepton \pt\ spectra are also sensitive to the Higgs mass, while details of the QCD modelling of the Higgs \pt\ spectrum are not important for the Higgs mass measurement.
 
It is found that for the envisaged mass accuracy of about $\pm$ 2 to 2.5 GeV, the lepton \pt\ spectra can be considered as  
essentially independent from details of the QCD modelling of the Higgs \pt\ spectrum. 

Combining the hypothetical cross section measurement with the lepton \pt\ spectra
and the estimated systematic uncertainties of about 10-15\%, associated with this 
signature, we find that the mass of a Standard-Model-Higgs signal in the mass range from 150-180 GeV 
can be measured with an accuracy of about 2-2.5 GeV. In case that no further improvements 
in the systematics for this channel can be achieved, our analysis  
shows also that such a mass measurement will be dominated by 
systematic uncertainties once integrated luminosities of about 10-20 fb$^{-1}$   
can be analysed.  

\section{acknowledgements}
We would like to thank Anne-Sylvie Giolo-Nicollerat for valuable comments and the very nice collaboration on the systematics and discovery potential.


\begin{thebibliography}{10}


\bibitem{search}
ATLAS collaboration, ``Atlas detector and physics performance; 
technical design report.'', vol. 2, report CERN/LHCC 99-15, ATLAS-TDR-15;\\
CMS collaboration, ``CMS detector and physics performance; 
technical design report.'', vol. 2, report CERN/LHCC.


\bibitem{jakobs}
  V.~Buscher and K.~Jakobs,
  Int.\ J.\ Mod.\ Phys.\ A {\bf 20}, 2523 (2005)
  [arXiv:hep-ph/0504099].

\bibitem{djouadi_rev}
  A.~Djouadi,
  arXiv:hep-ph/0503172.


\bibitem{carena}
M.~Carena and H.~E.~Haber,
Prog.\ Part.\ Nucl.\ Phys.\  {\bf 50}, 63 (2003)
[arXiv:hep-ph/0208209].

\bibitem{Dittmar:1996ss}
M.~Dittmar and H.~K.~Dreiner,
Phys.\ Rev.\ D {\bf 55}, 167 (1997)
[arXiv:hep-ph/9608317];


\bibitem{genzinski}
D. Genzinski, 
$\rm{http://ichep06.jinr.ru/reports/17\_15p10\_Glenzinski.ppt}$
and further references therein.




\bibitem{dittmareps97}
M.~Dittmar and H.~K.~Dreiner, Contributed Paper to the 1997 EPS conference in Jerusalem 
and CMS Note 97/080.

\bibitem{Glover:1988fn}
  E.~W.~N.~Glover, J.~Ohnemus and S.~S.~D.~Willenbrock,
  Phys.\ Rev.\ D {\bf 37} (1988) 3193.


\bibitem{Rainwater:1999sd}
  D.~L.~Rainwater and D.~Zeppenfeld,
  Phys.\ Rev.\  D {\bf 60} (1999) 113004
  [Erratum-ibid.\  D {\bf 61} (2000) 099901]
  [arXiv:hep-ph/9906218].




\bibitem{Davatz:2004zg}
  G.~Davatz, G.~Dissertori, M.~Dittmar, M.~Grazzini and F.~Pauss,
  JHEP {\bf 0405} (2004) 009
  [arXiv:hep-ph/0402218].

\bibitem{Davatz:2006ut}
  G.~Davatz, F.~Stockli, C.~Anastasiou, G.~Dissertori, M.~Dittmar, K.~Melnikov and F.~Petriello,
  JHEP {\bf 0607} (2006) 037
  [arXiv:hep-ph/0604077].

\bibitem{Anastasiou:2005qj}
  C.~Anastasiou, K.~Melnikov and F.~Petriello,
  Nucl.\ Phys.\ B {\bf 724} (2005) 197
  [arXiv:hep-ph/0501130].



\bibitem{Davatz:2006kb}
  G.~Davatz, M.~Dittmar and A.~S.~Giolo-Nicollerat,
CERN-CMS-NOTE-2006-047



\bibitem{Sjostrand:2006za}
  T.~Sjostrand, S.~Mrenna and P.~Skands,
  JHEP {\bf 0605} (2006) 026
  [arXiv:hep-ph/0603175].

\bibitem{Bozzi:2003jy}
  G.~Bozzi, S.~Catani, D.~de Florian and M.~Grazzini,
  Phys.\ Lett.\ B {\bf 564}, 65 (2003)
  [arXiv:hep-ph/0302104].

\bibitem{Bozzi:2005wk}
  G.~Bozzi, S.~Catani, D.~de Florian and M.~Grazzini,
  Nucl.\ Phys.\ B {\bf 737}, 73 (2006)
  [arXiv:hep-ph/0508068].

\bibitem{Sullivan:2006hb}
  Z.~Sullivan and E.~L.~Berger,
  Phys.\ Rev.\ D {\bf 74} (2006) 033008
  [arXiv:hep-ph/0606271].

\bibitem{Slabospitsky:2002ag}
  S.~R.~Slabospitsky and L.~Sonnenschein,
  Comput.\ Phys.\ Commun.\  {\bf 148} (2002) 87
  [arXiv:hep-ph/0201292].



\bibitem{Binoth:2005ua}
  T.~Binoth, M.~Ciccolini, N.~Kauer and M.~Kramer,
  JHEP {\bf 0503} (2005) 065
  [arXiv:hep-ph/0503094].

\bibitem{D'Hondt:2006ds}
  J.~D'Hondt, J.~Heyninck and S.~Lowette,
CERN-CMS-NOTE-2006-064

\bibitem{Moch:2005ky}
  S.~Moch and A.~Vogt,
  Phys.\ Lett.\ B {\bf 631} (2005) 48
  [arXiv:hep-ph/0508265].
 \bibitem{Dittmar:1997md}
 M.~Dittmar, F.~Pauss and D.~Zurcher,
 Phys.\ Rev.\ D {\bf 56} (1997) 7284
 [arXiv:hep-ex/9705004].


\bibitem{Martin:2002aw}
  A.~D.~Martin, R.~G.~Roberts, W.~J.~Stirling and R.~S.~Thorne,
  Eur.\ Phys.\ J.\ C {\bf 28} (2003) 455
  [arXiv:hep-ph/0211080].




\bibitem{Corcella:2000bw}
  G.~Corcella {\it et al.},
  JHEP {\bf 0101} (2001) 010
  [arXiv:hep-ph/0011363].

\bibitem{Frixione:2003ei}
  S.~Frixione, P.~Nason and B.~R.~Webber,
  JHEP {\bf 0308} (2003) 007
  [arXiv:hep-ph/0305252].
  S.~Frixione and B.~R.~Webber,
  JHEP {\bf 0206} (2002) 029
  [arXiv:hep-ph/0204244].



\bibitem{Jung:2001hx}
  H.~Jung,
  Comput.\ Phys.\ Commun.\  {\bf 143} (2002) 100
  [arXiv:hep-ph/0109102].

\bibitem{Alekhin:2005dx}
  S.~Alekhin {\it et al.},
   ``HERA and the LHC - A workshop on the implications of HERA for LHC  physics:
  arXiv:hep-ph/0601012.

\bibitem{Buttar:2006zd}
  C.~Buttar {\it et al.},
   ``Les Houches physics at TeV colliders 2005, standard model, QCD, EW, and
  arXiv:hep-ph/0604120.




\end{thebibliography}

\end{document}